\documentclass[final,3p,times]{elsarticle}
\usepackage{amssymb}
\usepackage{amsmath}
\usepackage{multicol}
\usepackage[english]{babel}
\usepackage{color}
\usepackage{bm}

\newcommand{\dd}{\text{d}}
\newcommand{\ee}{\text{e}}

\newcommand{\p}{\partial}

\providecommand{\avg}[1]{\left \langle #1 \right \rangle}
\providecommand{\avgg}[1]{\big \langle #1 \big \rangle}
\providecommand{\pnt}[1]{\left ( #1 \right)}
\providecommand{\brt}[1]{\left [ #1 \right]}
\providecommand{\brtg}[1]{\big [ #1 \big]}

\providecommand{\abs}[1]{\left| #1 \right|}

\providecommand{\f}[2]{\frac{#1}{#2}}
\providecommand{\df}[2]{\dfrac{#1}{#2}}

\journal{Physica A}

\begin{document}

\begin{frontmatter}

\title{The statistical physics of active matter:\\from self-catalytic colloids to living cells}

\author[label1]{\'Etienne Fodor}
\author[label2]{M. Cristina Marchetti}

\address[label1]{DAMTP, Centre for Mathematical Sciences, University of Cambridge, Wilberforce Road, Cambridge CB3 0WA, United Kingdom}
\address[label2]{Physics Department and Soft Matter Program, Syracuse University, Syracuse, NY 13244, USA}

\begin{abstract}

These lecture notes are designed to provide a brief introduction into the phenomenology of active matter and to present some of the analytical tools used to rationalize the emergent behavior of active systems. Such systems are made of interacting agents able to extract energy stored in the environment to produce sustained directed motion. The local conversion of energy into mechanical work drives the system far from equilibrium, yielding new dynamics and phases. The emerging phenomena can be classified depending on the symmetry of the active particles and on the type of microscopic interactions. We focus here on steric and aligning interactions, as well as interactions driven by shape changes. The models that we present are all inspired by experimental realizations of either synthetic, biomimetic or living systems. Based on minimal ingredients, they are meant to bring a simple and synthetic understanding of the complex phenomenology of active matter.

\end{abstract}


\end{frontmatter}


\section{Introduction}

Active matter consists of systems made of a large number of interacting constituents able to convert some source of energy stored in the environment into directed motion~\cite{Ramaswamy2010, Marchetti2013, Bechinger2016}. In contrast with driven systems, for which the departure from equilibrium is controlled by an external field or through boundary conditions, the breakdown of time reversal symmetry, characteristic of nonequilibrium dynamics, occurs at the level of individual components. The drive is local and sustained, being independent for each active agent. The large scale behavior emerges from collective self-organization, leading to novel phenomena such as nonequilibrium phase transitions and collective directed motion.

Examples of interacting self-propelled agents can be found at different scales. Bacteria and self-catalytic colloids are canonical examples of active particles at the micro-scale, with typical self-propulsion speed of the order of $10\,\mu\text{m}/s$~\cite{Dombrowski2004, Sokolov2007, Palacci2013}. Tracers in living cells, reflecting the intracellular activity of molecular motors and cytoskeletal filaments, also follow an active dynamics~\cite{Guo2014, Guo2015, Turlier2016, Ahmed2016}. A living cell in itself can be regarded as an active ``particle'', with size of about $10\,\mu\text{m}$ and speed of the order of $10\,\mu\text{m}/\text{hour}$. Assembly of such cells in epithelial tissues exhibit collective migration that underlies development and organ formation~\cite{Gorfinkiel2011, Guillot2013, Bi2016}. At larger scales, groups of animals such as bird flocks, fish schools, or even a human crowd can be modeled as interacting self-propelled agents~\cite{Cavagna2014}.

As a novel class of nonequilibrium systems, active matter has been at the center of various studies during the past decades. The modeling of active systems merges tools from statistical mechanics, soft matter and hydrodynamics. One of the main goal is to explore and classify  the various  emergent phases, and to understand how phase transitions are controlled by the microscopic interactions~\cite{Marchetti2013}. Identifying generic properties of such phases allows one to classify the spontaneous self-organization depending on the type of active components. As an example, the orientational order depend on the shape and symmetry of the microscopic agents: (i)  ferromagnetic-like order is observed when the particles are polar, namely when they have distinct head and tail~\cite{Vicsek1995, Toner1995}; (ii)  nematic order is reported for apolar particles with head-tail symmetry~\cite{Chate2006, Schaller2010}; (iii) no orientational order appears when particles are spherical in shape, yet other surprising collective behaviors emerges due to the self-propulsion~\cite{Fily2012, Cates2015, Bechinger2016}. The role of the surrounding medium allows for another classification. We focus here on the case where the medium is regarded as an inert substrate only providing some passive friction, so that the momentum of the particles is not conserved.

The lecture notes are organized as follows. After a brief review of the dynamics of a passive Brownian particle in an equilibrium thermal bath, Sec.~\ref{sec:single} presents  generic minimal models that have emerged as paradigm for the phenomenological modeling of collections of self-propelled particles. We first describe noninteracting particles by discussing the properties of both the free motion and the density profile under confinement. Section~\ref{sec:inter} is dedicated to interacting agents. We first discuss strategies that can be employed to obtain an effective hydrodynamic theory. We present separately the treatment of steric and alignment interactions, revealing respectively the possibility of a phase separation even when interactions are purely repulsive, and the emergence of collective directed motion in ordered states. Finally, we discuss how particle models have recently been adapted to describe dense epithelial tissues by merging the Vertex Model that describes cells as irregular polygons that tile the plane with active matter ideas to develop a model of motile cells where the behavior is tuned by cellular shape. We conclude with a brief discussion and outlook.


\section{A single active particle: a first insight into the phenomenology of active matter}\label{sec:single}

\subsection{Passive Brownian particle}

In the early twentieth century, the experiment by Jean Perrin was one of the first attempts to extract  quantitative information from the random trajectories of colloidal grains suspended in water~\cite{Perrin}. Perrin noticed that the instantaneous velocity of the grains could not be quantified properly due to the discontinuities in their trajectories. This finding motivated the phenomenological description proposed by Paul Langevin~\cite{Langevin}. He postulated that the effect of the solvent on the colloids could be separated into two contributions: a mean drag force $ - \zeta {\bf v} $ opposed to the displacement, where $\zeta$ denotes the friction coefficient, and a random force $(2 B)^{1/2} \boldsymbol \xi$ describing the effect of collisions by solvents atoms that drive the colloid motion. Assuming that the colloid has a mass $m$ and is subject to a potential $U$, the dynamics follows from Newton's second law,
\begin{equation}\label{eq:dyn_pbp}
	m \dot {\bf v} = - \nabla U - \zeta {\bf v} + (2B)^{1/2} {\boldsymbol \xi} .
\end{equation}
This is the seminal Langevin equation, in its underdamped version, describing the dynamics of a passive Brownian particle (PBP). The random force is taken as a Gaussian white noise with correlations $ \avgg{ \xi_\alpha (t) \xi_\beta (0) } = \delta_{\alpha\beta} \delta (t) $. The Langevin equation allows one to predict the time evolution of a number of observables. In the absence of potential, the average velocity squared is given by
\begin{equation}
	\avgg{ \brt{ {\bf v} (t) }^2 } = d \f{B}{m \zeta} \pnt{ 1 - \ee^{ - 2 \abs t / \tau_\text{m} } } ,
\end{equation}
where we have introduced the inertial time scale $\tau_\text{m} = m / \zeta$, and $d$ denotes the spatial dimension. We have assumed that the initial velocity is zero. At large times, the equipartition theorem relates the velocity fluctuations to the solvent temperature as $ \avgg{ {\bf v}^2 } = d (T / m) $, where we have set the Boltzmann constant to unity. The amplitude of the noise is then fixed by $ B = \zeta T $. In the absence of potential, the mean-square displacement (MSD) $ \avgg{ \Delta {\bf r}^2 (t) } = \avgg{ \brt{ {\bf r} (t) - {\bf r} (0) }^2 } $ can be obtained from Eq.~\eqref{eq:dyn_pbp} as
\begin{equation}\label{eq:msd_pbp}
	\avgg{ \Delta {\bf r}^2 (t) } =2 d  \f{T}{\zeta} \brt{ \abs{t} - \tau_\text{m} \pnt{ 1 - \ee^{ - \abs t / \tau_\text{m} } } } =
	\begin{cases}
		d\df{T}{\zeta} \df{t^2}{\tau_\text{m}} \quad \text{for} \,\, t\ll\tau_\text{m} ,
		\\\\
		2 d \df{T}{\zeta} \abs t \quad \text{for} \,\, t\gg\tau_\text{m}.
	\end{cases}
\end{equation}
The particle's motion is ballistic at short times and diffusive at large times. The translational diffusion coefficient $ D_\text{t} = {\lim}_{t\to\infty} \avgg{ \Delta {\bf r}^2 (t) } / 2 d t $ can be expressed in terms of the temperature and the friction coefficient via the Einstein relation: $ D_\text{t} = T / \zeta $. The diffusion coefficient is also related to the velocity autocorrelation through the Green-Kubo formula,
\begin{equation}
	D_\text{t} = \f{1}{d} \int_0^\infty \avg{ {\bf v} (t) \cdot {\bf v} (0) } \dd t .
\end{equation}
This can be regarded as a simplified statement of the fluctuation-dissipation theorem which connects the amplitude of fluctuations with the relaxation of the system~\cite{Kubo1966}. It formally expresses the fact that the random force and the damping force originate from the same microscopic processes, namely the collision between the colloid and the surrounding solvent particles.


\subsection{Self-propelled particle}

To rationalize the emergent phenomena observed in experimental systems of self-propelled particles, minimal models have been proposed for the self-propulsion~\cite{Bechinger2016}. This is implemented by adding to the particle's dynamics a sustained energy source that embodies the microscopic conversion of energy stored in the environment into a directed motion. The resulting propulsion force is subject to  fluctuations which are generally not thermal in origin. Provided that inertial effects can be neglected, the dynamics is given by a force balance between the damping force $ - \zeta \dot {\bf r} $, the self-propulsion $\bf u$ and forces deriving from a potential $- \bm\nabla U$:
\begin{equation}\label{eq:dyn_spp}
	\dot {\bf r} = {\bf u} - \mu_\text{t} \bm\nabla U + (2 D_\text{t})^{1/2} {\boldsymbol\xi} ,
\end{equation}
where $\mu_\text{t} = \zeta^{-1}$ denotes the translational mobility. The directionality of the self-propulsion, namely the ability of the orientation to stay constant during a typical persistence time $\tau$, is captured by assuming that its correlations decay exponentially in time,
\begin{equation}\label{eq:corr}
	\avgg{ u_\alpha (t) u_\beta (0) } = \delta_{\alpha\beta} \f{v_0^2}{d} \ee^{ - \abs t / \tau } .
\end{equation}
Various models of self-propulsion differ by the assumptions made on the higher order cumulants of the statistics. We present below three of such models: run-and-tumble particles (RTPs), active Brownian particles (ABPs) and active Ornstein-Uhlenbeck particles (AOUPs).

\begin{figure}
		\centering
		\includegraphics[width=0.85\columnwidth]{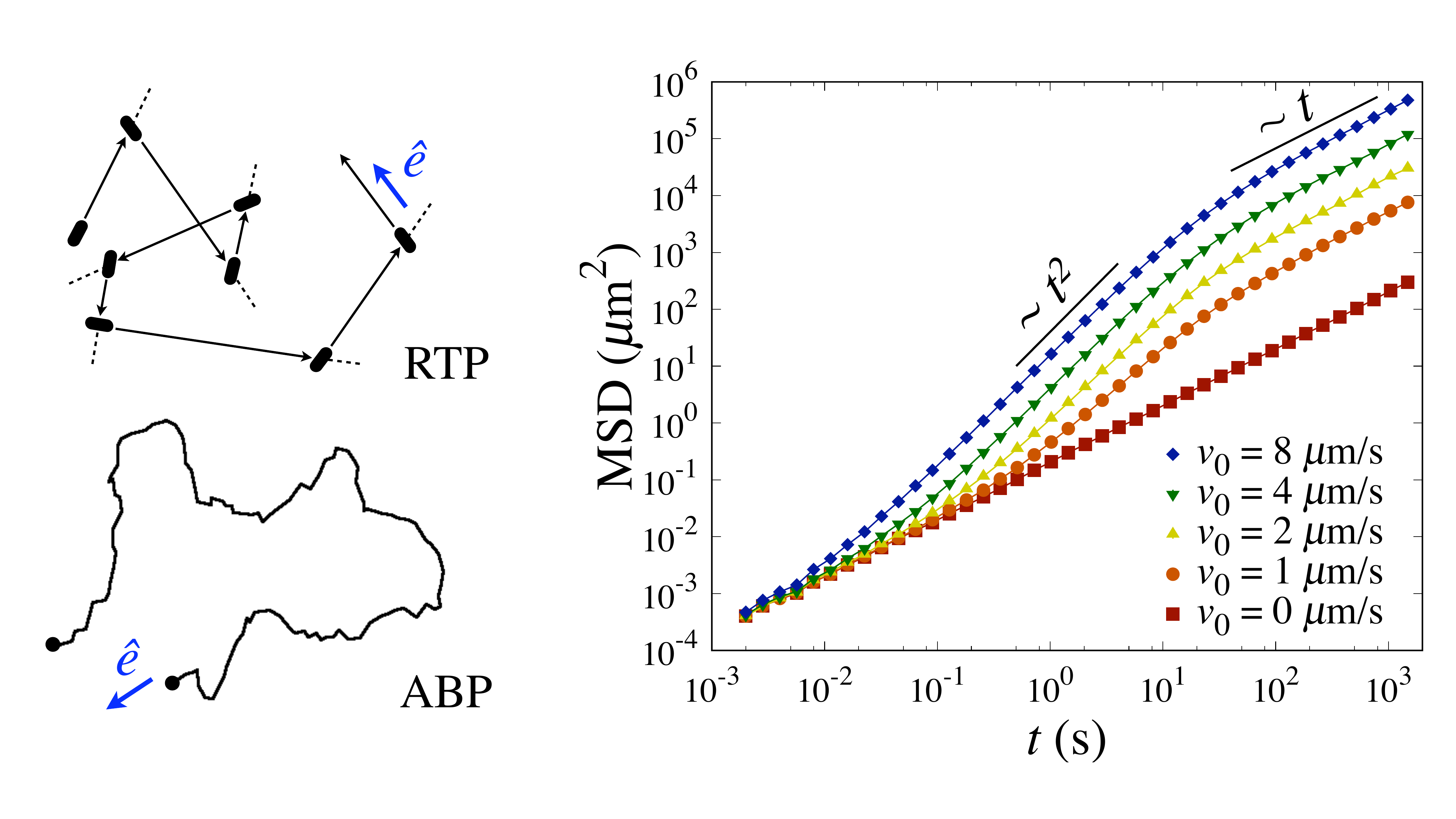}
	\caption{\label{fig:traj}
		Typical trajectory of (Top left) a run-and-tumble particle with stochastic instantaneous changes of the self-propulsion orientation, and (Bottom left) an active Brownian particle for which the orientation decorrelates through angular diffusion. 
		(Right) Mean-square displacement (MSD) as a function of time for non-interacting ABPs with $D_\text{t}=0.1 \, \mu\text{m}^2.\text{s}$ and $D_\text{r} = 0.1 \,\text{s}^{-1}$.
	}
\end{figure}

The run-and-tumble motion is inspired by the dynamics of bacteria~\cite{Schnitzer1993, Tailleur2008}. It alternates between an active state when the particles moves at constant speed $v_0$ in a given direction (``run''), and a passive state when the center of mass of the particle stays constant while reorienting its direction (``tumble''). In practice, the change of direction is taken as instantaneous and completely isotropic, occurring with a given rate $\nu$, so that typical trajectories are made of straight lines with random length of typical size $v_0 / \nu$, as shown in Fig.~\ref{fig:traj}. ABPs were introduced to mimick the dynamics of self-phoretic colloids with asymmetric chemical and/or physical properties~\cite{Fily2012, Hagan2013}. The self-propulsion has a fixed norm as for RTPs, yet the angular direction  is now controlled by a diffusive process. In two dimensions, the self-propulsion is written as $ {\bf u} = v_0 ( \cos \theta, \sin \theta) $ with angular dynamics
\begin{equation}
	\dot \theta = (2 D_\text{r})^{1/2} \eta ,
\end{equation}
where $\eta$ is a Gaussian noise with correlations $ \avg{ \eta(t) \eta(0) } = \delta(t) $, and $D_\text{r}$ denotes the rotational diffusion coefficient. Considering a self-propelled particle whose change in direction can occur due to both instantaneous reorientation and angular diffusion, the correlations of self-propulsion are given by~\eqref{eq:corr} with $\tau^{-1} = \nu + (d-1) D_\text{r} $~\cite{Tailleur2013}: this is the relation between the persistence of the self-propulsion fluctuations and the self-propulsion angular relaxation for RTPs and ABPs. The full statistics is not Gaussian, with different higher order cumulants for RTPs and ABPs.

The dynamics of AOUPs was originally proposed as an approximated treatment of ABPs~\cite{Maggi2015, Farage2015}. It consists in neglecting the non-Gaussian nature of the self-propulsion statistics, which amounts to allowing the amplitude of self-propulsion to fluctuate. In practice, such an approximation captures the emergent behavior of ABPs with a better accuracy in three dimensions than two. The self-propulsion dynamics can be described as an Ornstein-Uhlenbeck process
\begin{equation}
	\tau \dot {\bf u} = - {\bf u} + (2 D_\text{ac})^{1/2} {\boldsymbol\chi} ,
\end{equation}
where we have introduced an active diffusion coefficient $D_\text{ac} = \mu_\text{t} v_0^2 \tau / d$, and $\boldsymbol\chi$ is a Gaussian noise with correlations $ \avgg{ \chi_\alpha (t) \chi_\beta (0) } = \delta_{\alpha\beta} \delta(t) $ completely uncorrelated with $\boldsymbol\xi$. It follows that the self-propulsion correlations are indeed given by~\eqref{eq:corr}. When describing the dynamics and structure of AOUPs, the translational noise can be neglected to proceed within some specific approximations,\footnote{For a micron-size colloid suspended in water at ambient temperature, the translational diffusion coefficient is of the order of $ D_\text{t} \simeq 0.1 \, \mu\text{m}^2 \,.\, \text{s}^{-1}$, whereas the active diffusion coefficient of self-phoretic colloids is of the order of $ D_\text{ac} \simeq 10^2	\, \mu\text{m}^2 \,.\, \text{s}^{-1}$.} in which case the dynamics can be written as
\begin{equation}\label{eq:dyn_aoup}
	\tau \dot {\bf v} = - {\bf v} - \mu_\text{t} ( 1 + \tau {\bf v} \cdot \nabla ) \nabla U + (2 D_\text{ac})^{1/2} {\boldsymbol\chi} .
\end{equation}
Comparing this to the dynamics of an underdamped PBP given in Eq.~\eqref{eq:dyn_pbp}, we see that the effect of the noise persistence amounts to  introducing (i) an effective inertia controlled by the persistence time $\tau$ and (ii) a velocity-dependent mobility. Moreover, in the limit of vanishing persistence at fixed $D_\text{ac}$ the original dynamics given in Eq.~\eqref{eq:dyn_spp} reduces to the one of an overdamped PBP at a temperature $ \zeta ( D_\text{t} + D_\text{ac} ) $. Therefore, the persistence can be regarded as the only parameter monitoring the nonequilibrium properties of an AOUP~\cite{Nardini2016}.


\subsubsection*{Free motion: how different from a passive Brownian particle?}

To explore the properties of a self-propelled particle, let us first consider the case of an isolated particle. The MSD is identical for the three models of self-propulsion presented above, when both translation and rotational noises are considered, since it only depends on the two-point correlations of the fluctuations. It can be deduced from~(\ref{eq:dyn_spp}--\ref{eq:corr}) as
\begin{equation}
	\avgg{ \Delta {\bf r}^2 (t) } = 2 d \pnt{ D_\text{t} + D_\text{ac} } \abs t + 2 ( v_0 \tau )^2 \pnt{ \ee^{ - \abs t / \tau } - 1 } =
	\begin{cases}
		2 d D_\text{t} \abs t \quad \text{for} \,\, t\ll\tau  ,
		\\\\
		2 d \pnt{ D_\text{t} + D_\text{ac} } \abs t \quad \text{for} \,\, t\gg\tau .
	\end{cases}
\end{equation}
An intermediate ballistic regime appears for large self-propulsion amplitude $ \avgg{ \Delta {\bf r}^2 (t) } = ( v_0 t )^2 $, qualitatively similar to the case of an underdamped PBP in~\eqref{eq:msd_pbp}. The particle has a diffusion behavior at short and large times with different diffusion coefficients, as shown in Fig.~\eqref{fig:traj}. Therefore, the dynamics of an isolated self-propelled particle at times and distances larger than, respectively, the persistence time $\tau$ and the persistence length $v_0 \tau$ can not be distinguished from a ``hot'' PBP at temperature $ \zeta ( D_\text{t} + D_\text{ac} ) $.


\subsubsection*{Profile under confinement: accumulation at the boundaries}

To go beyond the simple case of an isolated  self-propelled particle, we now consider that an external confining potential is applied to the particle. The steady state is not known in general for any of the three models of self-propulsion presented above, with the notable exceptions of RTPs in one dimension and AOUPs in a harmonic trap. Various approximation schemes have been proposed to capture asymptotic behaviors of the distribution. After stating the few exact results available, we discuss below some of the most successful attempts to get approximated ones.

\begin{figure}
	\begin{multicols}{2}
		\centering
		\includegraphics[width=.78\columnwidth]{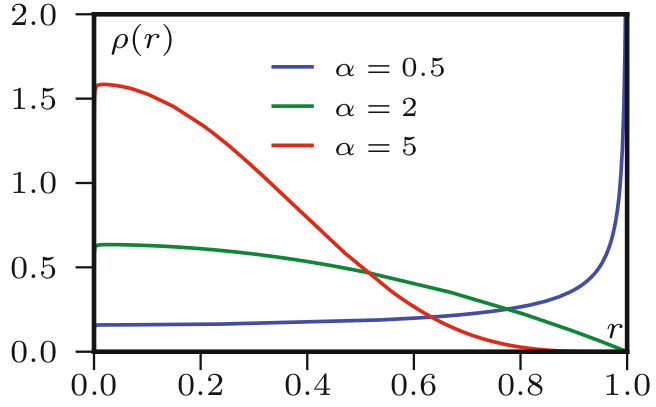}
		\vskip.5cm
		\includegraphics[width=.8\columnwidth]{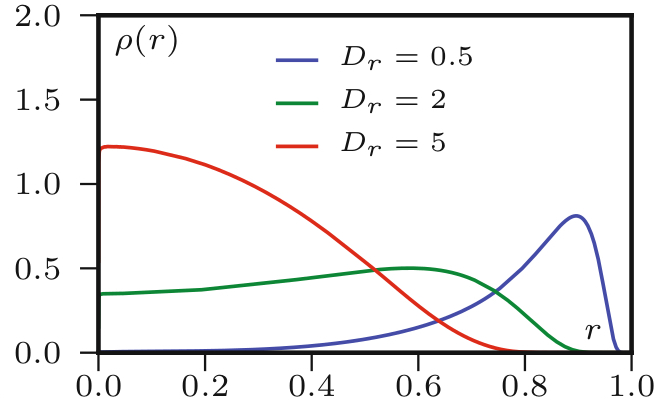}

		\columnbreak
		\centering
		~
		\vskip1cm
		\includegraphics[width=.9\columnwidth]{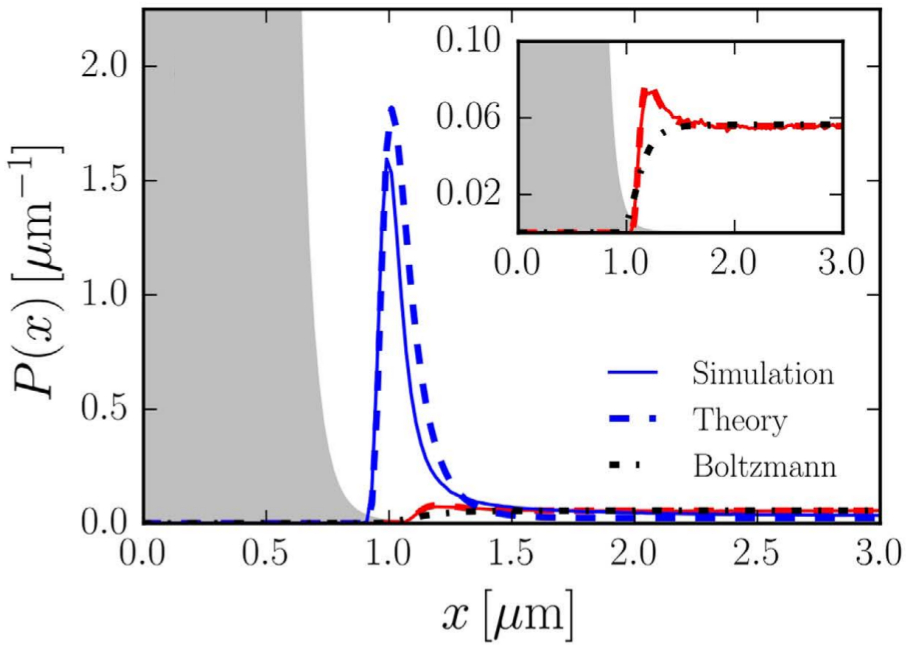}
	\end{multicols}
	\caption{\label{fig:profile}
		Stationary distribution of non-interacting self-propelled particles.
		Profile under confinement by a harmonic trap in two dimensions as a function of the distance to the trap centre, obtained from numerical simulations of (Top left) RTPs and (Bottom left) ABPs. Taken from~\cite{Solon2015c}.
		(Right) Profile of non-interacting APBs close to an external wall of the form $U(x) = x^{-12}$ in one dimension, with $D_\text{t} = 0$. Higher peak: $D_\text{ac} = 100 \, \mu\text{m}^2/\text{s}$, $\tau = 1 \, \text{s}$. Lower peak: $D_\text{ac} = 0.4 \, \mu\text{m}^2/\text{s}$, $\tau = 0.1 \, \text{s}$, zoom in the inset. The dashed line corresponds to the prediction of UCNA. Taken from~\cite{Maggi2015}.
	}
\end{figure}

Considering RTPs in one dimension subjected to a potential $U$, the stationary distribution reads~\cite{Bechinger2016}
\begin{equation}
	p(x) = \f{p_0}{ 1 - ( \mu_\text{t} \p_x U / v_0 )^2  } \exp \brt{ - \f{ \mu_\text{t} }{v_0^2 \tau} \int_0^x \f{ \p_y U \dd y }{ 1 - ( \mu_\text{t} \p_y U / v_0 )^2 } } ,
\end{equation}
where $p_0$ is a normalization factor. In the case of harmonic potential of the form $ U = \kappa x^2 / 2 $, the particles are confined in a region $ \abs x < v_0 / \mu_\text{t} \kappa $ and the distribution can be simplified as
\begin{equation}
	p(x) = p_0 \brt{ 1 - ( \mu_\text{t} \kappa x / v_0 )^2 }^{ 1 / 2 \mu_\text{t} \kappa \tau - 1 } .
\end{equation}
The distribution is bimodal when $2 \mu_\text{t} \kappa \tau > 1$ with accumulation of particles at a distance $ v_0 / \mu_\text{t} \kappa $ from the trap centre. Such an accumulation is a generic feature of self-propelled particles for any type of confining potential. The case of AOUPs in a harmonic potential has a special status for any dimension due to the mapping into an underdamped dynamics when the translational noise is neglected: the AOUP dynamics~\eqref{eq:dyn_aoup} is similar with the PBP one~\eqref{eq:dyn_pbp} under the transformation $ \tau / ( 1 + \mu_\text{t} \kappa \tau ) \to \tau_\text{m} $, $ \kappa / ( 1 + \mu_\text{t} \kappa \tau) \to \kappa $ and $ D_\text{ac} / ( 1 + \mu_\text{t} \kappa \tau)^2 \to D_\text{t} $. It follows that the steady state is given by a Boltzmann-like distribution as $ p ({\bf r}) = p_0 \exp \brt{ - ( 1 + \mu_\text{t} \kappa \tau ) \kappa {\bf r}^2 / 2 \zeta D_\text{ac} } $~\cite{Szamel2014}. For RTPs and ABPs, a Boltzmann-like factor of the form $ p({\bf r}) = p_0 \exp \brt{ - \kappa {\bf r}^2 / 2 \zeta (D_\text{t} + D_\text{ac}) } $ is also recovered for any dimension in the regime where the persistence length $ v_0 \tau $ is smaller than the typical trap radius $ v_0 / \mu_\text{t} \kappa $, namely when $ \mu_\text{t} \kappa \tau \ll 1$. Otherwise, accumulation is observed at the boundaries of the potential~\cite{Solon2015c}, as reported in Fig.~\ref{fig:profile}.

Considering AOUPs in a generic potential when neglecting translational noise, two main approximation schemes have been proposed to obtain the stationary distribution. One relies on an approximated equivalent dynamics referred to as the Unified Colored-Noise Approximation (UCNA), the other is a systematic perturbation of the steady state. The UCNA amounts to neglecting the inertial-like effects on the dynamics~\eqref{eq:dyn_aoup}. It reduces the dynamics into an overdamped equilibrium one for which the steady state can be determined exactly~\cite{Maggi2015, Marconi2015}:
\begin{equation}\label{eq:ucna}
	p ({\bf r}) = p_0 \abs{ \det {\mathbb M} } \exp \brt{ - \f{\mu_\text{t} U}{D_\text{ac}} - \f{ \tau \pnt{ \mu_\text{t} \nabla U }^2 }{2 D_\text{ac}} } ,
\end{equation}
where we have introduced the matrix $\mathbb M$ with components $ {\mathbb M}_{\alpha\beta} = \delta_{\alpha\beta} + \mu_\text{t} \tau \p^2_{\alpha
\beta} U $. Such a distribution has been shown to reproduce the profile of ABPs close to a wall for small values of the persistence~\cite{Maggi2015}, as shown in Fig.~\eqref{fig:profile}. Yet, provided that UCNA relies on an equilibrium mapping of the dynamics, it fails to capture the dynamical properties which are inherently out-of-equilibrium, such as the existence of a particle current in a ratchet. To describe properly nonequilibrium signatures of the dynamics, a recent work has proposed a systematic perturbation scheme in terms of the persistence~\cite{Nardini2016}. Introducing the scaled velocity as $ \bar {\bf v} = \tau^{1/2} {\bf v} $, the distribution in the phase space position-velocity can be determined as a functional of the potential up to second order in the perturbation:
\begin{equation}\label{eq:dist_aoup}
	\begin{aligned}
		P({\bf r}, \bar {\bf v}) = P_0 \ee^{ - ( \mu_\text{t} U + \bar {\bf v}^2 / 2 ) / D_\text{ac} } \bigg\{& 1 - \f{\tau \mu_\text{t}}{2 D_\text{ac}} \brt{ \mu_\text{t} ( \nabla U )^2 + ( \bar {\bf v} \cdot \nabla )^2 U - 3 D_\text{ac} \nabla^2 U }
		\\
		& + \f{\tau^{3/2} \mu_\text{t}}{6 D_\text{ac}} \brt{ ( \bar {\bf v} \cdot \nabla )^3 U - 3 D_\text{ac} ( \bar {\bf v} \cdot \nabla ) \nabla^2 U } + {\cal O} (\tau^2) \bigg\} .
	\end{aligned}
\end{equation}
Integrating over the velocities yields the position density profile of the form $ p ({\bf r}) = p_0 \ee^{ - \mu_\text{t} \tilde U / D_\text{ac} } $, where we have introduced an effective potential by analogy with the Boltzmann distribution as $ \tilde U = U - \tau \brt{ \mu_\text{t} (\nabla U)^2 / 2 - D_\text{ac} \nabla^2 U } + {\cal O} (\tau^2) $. Such a solution coincides with the UCNA distribution~\eqref{eq:ucna} at first order, showing that UCNA is exact at this order. Based on the full distribution~\eqref{eq:dist_aoup}, it has been shown that there exists a regime of small persistence where nonequilibrium signatures, such as the breakdown of time reversal invariance of the correlations, are hidden, and yet the steady-state statistics is non-Boltzmann, but rather given by the effective potential $\tilde U$~\cite{Nardini2016}.


\section{Structure and dynamics of interacting agents}\label{sec:inter}

In this section, we discuss how the nonequilibrium nature of the self-propulsion modifies the collective effects for interacting particles, yielding a rich and novel phenomenology with respect to passive particles. The complexity of the dynamics, inherent to the large number of degrees of freedom, can be reduced by using a hydrodynamic approach. It consists in a continuum description of the system at large scales in terms of a small number of coarse-grained fields. As a first step, one needs to identify the fields associated with some spontaneous broken symmetries and conservation laws. The corresponding effective theory can then be constructed from three different routes: (i) based on phenomenological arguments, such as the symmetries of the system~\cite{Toner1995, Simha2002}; (ii) by introducing constitutive equations between the fluxes and the forces identified from the entropy production rate close to equilibrium~\cite{Martin1972, Kruse2004}; (iii) \textit{via} explicit coarse-graining of the microscopic dynamics~\cite{Bertin2006, Peshkov2014}. The latter provides explicit expressions for the kinetic coefficients appearing in the theory, whereas such coefficients are unknown \textit{a priori} from the two other methods. Yet, the coarse-graining procedures generally need to be combined with some approximations to arrive at some closed forms for the dynamical equations.

We consider a set of $N$ ABPs in two dimensions able to interact both \textit{via} steric repulsion and alignment:
\begin{equation}\label{eq:dyn_inter}
	\dot {\bf r}_i = v_0 \hat {\bf e} (\theta_i) - \mu_\text{t} \nabla_i \sum_{j=1}^N V ( {\bf r}_j - {\bf r}_i ) + ( 2 D_\text{t} )^{1/2} {\boldsymbol\xi_i} ,
	\quad
	\dot \theta_i = \mu_\text{r} \sum_{j=1}^N {\cal T} ( \theta_j-\theta_i, {\bf r}_j - {\bf r}_i ) + ( 2 D_\text{r} )^{1/2} \eta_i ,
\end{equation}
where $\boldsymbol\xi$ and $\eta$ are uncorrelated Gaussian noises with respective correlations
\begin{equation}
	\avgg{ \xi_{i\alpha} (t) \xi_{j\beta} (0) } = \delta_{ij} \delta_{\alpha\beta} \delta (t)  ,
	\quad
	\avgg{ \eta_i (t) \eta_j (0) } = \delta_{ij} \delta(t) .
	\label{eq:noise}
\end{equation}
In practice, the repulsion and the alignment are taken as short range interactions, such as the pair-wise potential $V$ and torque ${\cal T}$ vanish when the distance between two particles exceeds a given radius. Introducing the microscopic density of position and orientation as
\begin{equation}\label{eq:psi}
	\psi ({\bf r}, \theta, t) = \sum_{i=1}^N \delta \brt{ {\bf r} - {\bf r}_i(t) } \delta \brt{ \theta - \theta_i(t) } ,
\end{equation}
the dynamics of $\psi$ follows from~\eqref{eq:dyn_inter} by using standard coarse-graining procedures {\it \`a la} Dean without relying on any approximation~\cite{Dean1996}:
\begin{equation}\label{eq:dyn_psi}
	\begin{aligned}
		\p_t \psi =& \, \nabla \cdot \brt{ \mu_\text{t} \psi( {\bf r}, \theta ) \int \psi( {\bf x}, \theta ) \nabla V ( {\bf r} - {\bf x} ) \dd {\bf x} - v_0 \hat {\bf e} (\theta) \psi + D_\text{t} \nabla \psi + ( 2 D_\text{t} \psi )^{1/2} {\boldsymbol\Lambda} }
		\\
		& + \p_\theta \brt{ - \mu_\text{r} \psi({\bf r}, \theta) \int \psi ({\bf x}, \phi) {\cal T}( \theta-\phi, {\bf r}-{\bf x} ) \dd \phi \dd {\bf x} + D_\text{r} \p_\theta \psi + ( 2 D_\text{r} \psi )^{1/2} \Gamma } ,
	\end{aligned}
\end{equation}
where $\boldsymbol\Lambda$ and $\Gamma$ are uncorrelated Gaussian noises with respective correlations
\begin{equation}
	\avgg{ \Lambda_\alpha ({\bf r}, t) \Lambda_\beta ({\bf x}, s) } = \delta_{\alpha\beta} \delta ( {\bf r} - {\bf x} ) \delta (t-s) ,
	\quad
	\avg{ \Gamma ({\bf r}, t) \Gamma ({\bf x}, s) } = \delta ({\bf r} - {\bf x}) \delta (t-s) .
\end{equation}
We stress that, while Eq.~\eqref{eq:dyn_psi} for the microscopic density field $\psi$ is exact in the sense that it contains the same amount of information as the microscopic one in~\eqref{eq:dyn_inter}, the equation for the average density $\langle\psi\rangle$ discussed below couples the one-particle density to the two-particle one through interactions.  In the followingn we use the simplest approximation and factorize the two-particle density as the product of two one-particle ones. This corresponds to neglecting  correlations, whose dynamics is ruled by a hierarchy of equations in a non-closed form. The treatment of such correlations is a hard task in general and remains an open challnege for active systems.


\subsection{Steric interactions: motility-induced phase separation and glassy dynamics}

\begin{figure}
	\centering
	\includegraphics[width=.275\columnwidth]{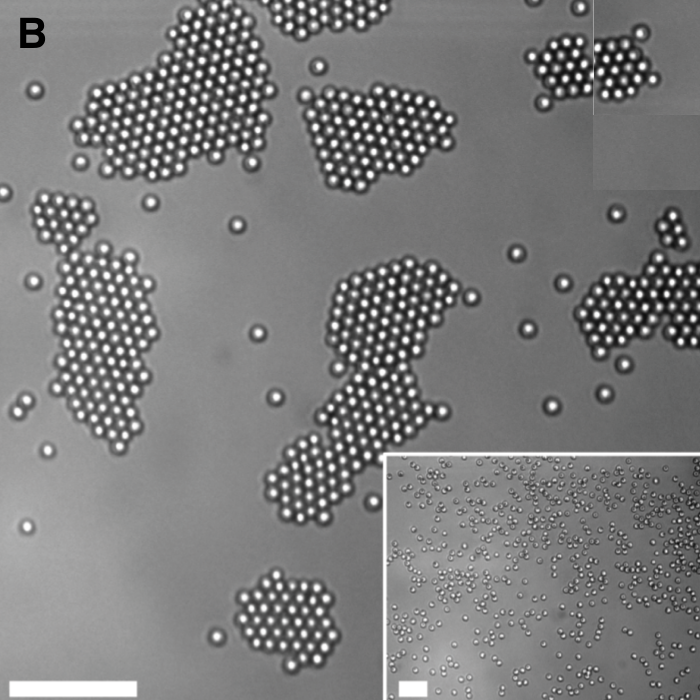}
	\hskip2cm
	\includegraphics[width=.28\columnwidth]{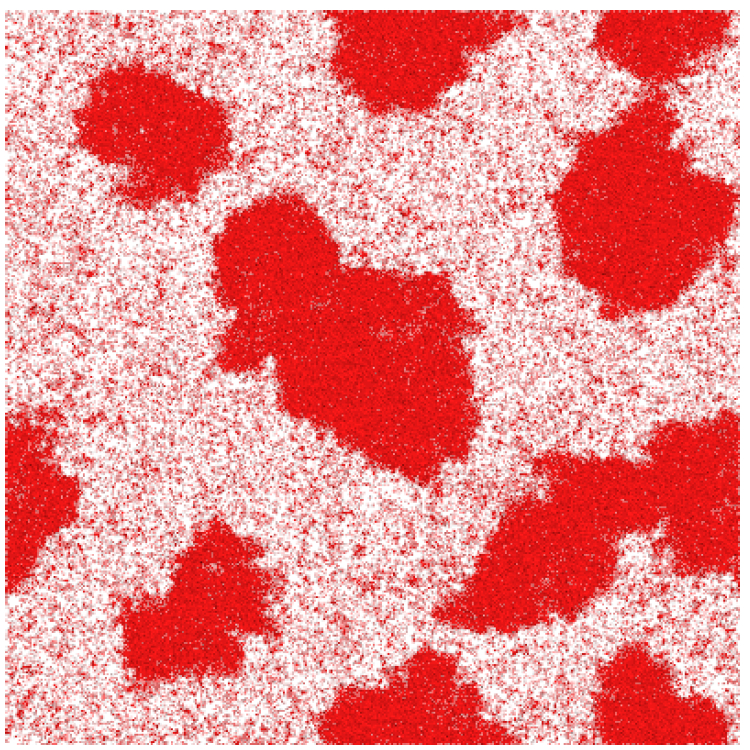}
	\caption{\label{fig:mips}
		(Left) Snapshot of self-phoretic colloids. The chemical reaction at the basis of the self-propulsion mechanism is controlled by the light shaded on the system, yielding the formation of clusters. The clusters disappear when the self-propulsion is stopped by turning the light off, as shown in the inset. Scale bar $10\,\mu\text{m}$. Taken from~\cite{Palacci2013}.
		(Right) Snapshot of some numerical simulations of interacting ABPs showing the emergence of large scale clusters. Taken from~\cite{Hagan2013}.
	}
\end{figure}

First, we consider the case where particles interact only through steric repulsion without any alignment. For passive Brownian particle, such  interactions leads to a homogeneous steady state whose structure is controlled by the density of particles. In contrast, the formation of clusters is observed when considering self-propelled particles, which can lead to a complete phase separation even though interactions are purely repulsive. This is the Motility-Induced Phase Separation (MIPS)~\cite{Tailleur2008,Fily2012,Cates2015}. The formation of clusters has been reported for different experimental realizations of self-phoretic colloids and swarms of bacteria~\cite{Buttinoni2013, Palacci2013}, as shown in Fig.~\ref{fig:mips}. The stability of distinct clusters, as opposed to complete phase separation, is often associated with the existence of long-range attractions, stemming for instance from chemotactic or hydrodynamic effects~\cite{Golestanian2012, Liebchen2015}. Moreover, arrested phase separation has been reported when the density is no longer conserved~\cite{Cates2010, Yang2014}.

A microscopic phenomenological picture has emerged to describe MIPS. It is based on the collisional slowing-down of the dynamics. Considering two particles colliding with opposite orientations, they will keep facing each other during a transient time associated with the persistence of self-propulsion. This results in an effective attraction that is at the basis of the formation of metastable clusters. Attractive effects are enhanced when the typical time between two successive collisions reduces, as controlled by self-propulsion speed and particle density. Such a picture has been rationalized through a mean-field theory at the hydrodynamic level that we present below.

The average density of particles is the hydrodynamic variable which controls the properties of the system at large scale. It is defined from the density of position and orientation~\eqref{eq:psi} by integrating over the angle and averaging ($\langle \cdot \rangle$) over the noise realizations:
\begin{equation}
	\rho ({\bf r}, t) = \int \avg{ \psi ({\bf r}, \theta, t) } \dd \theta .
\end{equation}
The dynamics of particle density can be obtained from the full dynamics~\eqref{eq:dyn_psi}. Integrating over the orientation yields the conservation of particle density, which can be written in the absence of interactions and translational diffusion as
\begin{equation}
	\p_t \rho = - v_0 \nabla \cdot {\bf p} ,
\end{equation}
where we have introduced the polarization density as
\begin{equation}
	{\bf p} ({\bf r}, t) = \int \hat {\bf e} (\theta) \avg{ \psi ({\bf r}, \theta, t) } \dd \theta .
\end{equation}
A hierarchy of equations for the higher-order angular moments can be deduced from~\eqref{eq:dyn_psi}. To close this hierarchy, we neglect all the moments beyond the polarization density. In the absence of interactions and translational diffusion, the polarization dynamics follows as
\begin{equation}
	\p_t {\bf p} = - D_\text{r} {\bf p} - \f{v_0}{2} \nabla \rho .
\end{equation}
At times larger than the relaxation time scale $ D_\text{r}^{-1} $, the polarization density can be eliminated to yield the following diffusion equation
\begin{equation}
	\p_t \rho = D_\text{ac} \nabla^2 \rho .
\end{equation}
The dynamics is similar to the one of PBPs with diffusion coefficient $D_\text{ac}$ in the absence of interactions and translational diffusion, as already observed from the free motion of self-propelled particles in such a case.

To account for the interactions, a Virial expansion valid at low density can be used. In such a limit, the effect of the interactions is to renormalize the self-propulsion speed as $ v_0 \to v(\rho) = v_0 ( 1 - \lambda \rho ) $, where $\lambda$ can be expressed in terms of the pair correlation function~\cite{Bialke2013, Fily2014}. The coupled dynamics of particle and polarization densities follows as
\begin{equation}
	\p_t \rho = - \nabla \cdot \brt{ v(\rho) {\bf p} } ,
	\quad
	\p_t {\bf p} = - D_\text{r} {\bf p} - \f{1}{2} \nabla \brt{ v(\rho) \rho } .
\end{equation}
Eliminating the polarization in the adiabatic limit yields
\begin{equation}
	\p_t \rho = \nabla \cdot \brt{ {\cal D}(\rho) \nabla \rho } ,
	\quad
	{\cal D}(\rho) = \f{\brt{ v(\rho) }^2 + \rho v(\rho) v'(\rho)}{2 D_\text{r}} ,
\end{equation}
where $ v'(\rho) = \dd v / \dd \rho < 0$. This effective mean-field dynamics predicts that spinodal decomposition occurs when the diffusivity $\cal D$ changes sign at high density. This supports the phenomenological picture that density homogeneities can build up from an effective reduction of the self-propulsion speed controlled by some crowding effects. A recent work has shown that, based an some explicit coarse-graining beyond mean-field, it is possible to reproduce quantitatively the phase diagram from a generalized Maxwell construction~\cite{Solon2016}. For RTPs, the steady state density is directly related to the effective self-propulsion speed as $ \rho_\text{s} \sim 1 / v (\rho_\text{s}) $. It follows that particles accumulate when they slow down. Such a self-trapping is enhanced by some feedback effect, yielding a spinodal decomposition similar to the one of ABPs~\cite{Tailleur2008, Cates2015}. Considering AOUPs, the spinodal lines can be predicted from liquid state theory on the basis of the equilibrium mapping provided by UCNA~\cite{Farage2015}. To describe nonequilibrium phase separation, alternative minimal models have also been proposed by postulating dynamical equations directly at the hydrodynamic level based on symmetry arguments~\cite{Wittkowski2014, Tiribocchi2015, Nardini2017}.

At even higher packing fraction, ABPs can form solid states, which exhibit crystalline order when the particles are monodispersed~\cite{Palacci2013, Reichhardt2014, Bialke2012, Menzel2014}, but are glassy in polydispersed systems~\cite{Berthier2014, Fily2014}. Active glasses form when the suppression of motility due to particle crowding is large enough to yield caged particles with effectively zero diffusivity. This density-driven glass transition has been quantified in simulations by examining the particles' mean square displacement that becomes bounded over the simulation time, indicating  an arrested state. It shares qualitative similarities with the glass transition observed at finite temperature in attractive colloids, where a glassy state is obtained by tuning the packing fraction above a critical value and by decreasing the temperature below a value where attractive forces overcome thermal fluctuations~\cite{Lu2013}. The onset of structural arrest in ABPs is controlled by (i) the packing fraction, (ii) the strength of self-propulsion $v_0$, and (iii) the persistence $\tau$ of the single particle dynamics. The self-propulsion speed plays a role naively similar to that of temperature, in the sense that the glass only occurs below a critical value of $v_0$. In general, however, the notion of effective temperature fails to describe the rich behavior of ABPs other than in very dilute or very dense limits~\cite{Levis2015}.
The dependence of the transition on the persistence of the single-particle dynamics is unique to active systems and is subtle in the limit of low $v_0$, where  increasing the persistence time as the glass transition is approached from the liquid side promotes large scale collective structural rearrangements that tend to fluidify the system, pushing the transition line to higher density than in thermal colloids~\cite{Fily2014, Berthier2014}. This can be understood by examining the low frequency or ``soft'' excitations of the jammed solid that are enhanced and become more collective in the limit $D_\text{r}\rightarrow 0$. Although the properties of the active glass resemble those of thermal colloidal glasses, the persistence of the dynamics provides a new knob that can be tuned to push the structural arrest into previously unexplored high density regions~\cite{Ni2014, Ni2014b}. To understand and predict the emergent behavior of active glasses, recent works have attempted to extend mode coupling theories, which have been used extensively to rationalize glassy dynamics in thermal systems, when self-propulsion comes at play~\cite{Szamel2015, Flenner2016, Nandi2017}.


\subsection{Alignment interactions: transition to collective directed motion}

\begin{figure}
	\centering
	\includegraphics[width=.425\columnwidth]{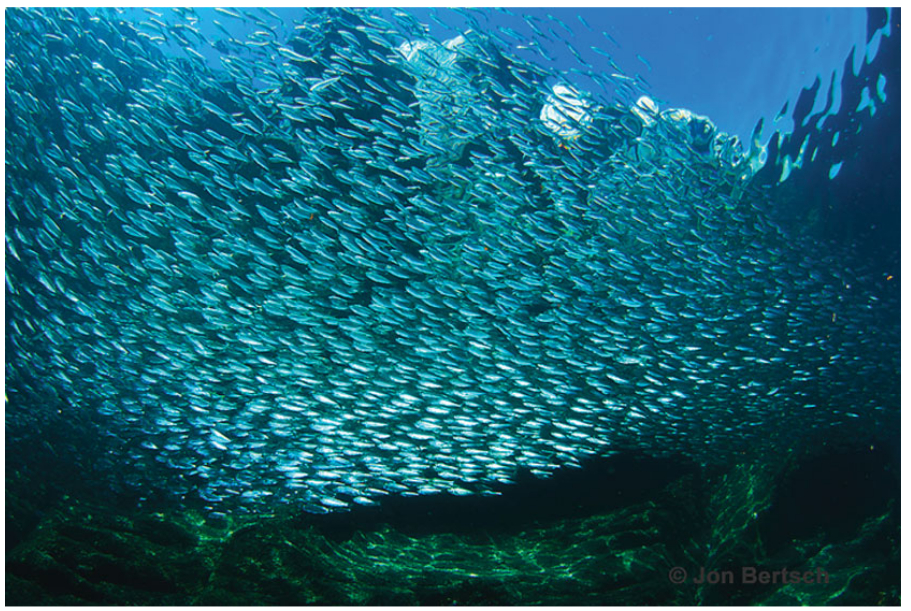}
	\hskip.5cm
	\includegraphics[width=.52\columnwidth]{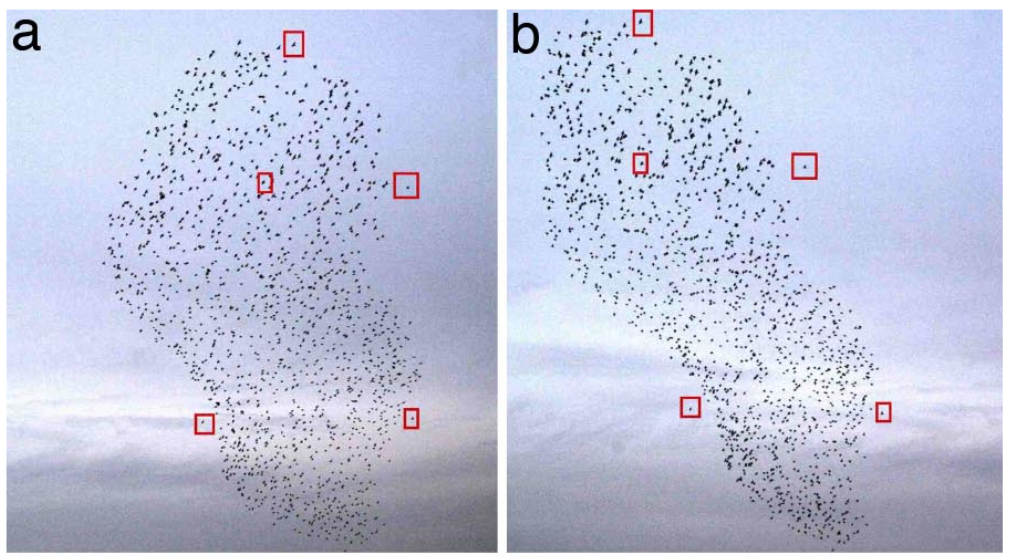}
	\caption{\label{fig:align}
		(Left) Image of a sardine school showing some polar order. Taken from~\cite{Marchetti2013}.
		(Middle-Right) Photographs of starlings taken as the same instant from two different locations. The red boxes indicate the same birds in each one of the images. Taken from~\cite{Ballerini2008}.
	}
\end{figure}

Self-propelled entities interact through alignment, in addition to their short-range repulsion. As an example, the steric hindrance between rod-like particles leads to an effective alignment~\cite{Narayan2007}. The collective motion observed in groups of animals, such as fish school or bird flocks for which no leader can be identified as shown in Fig.~\ref{fig:align}, is also often controlled by alignment interactions~\cite{Bialek2012, Cavagna2014}. A generic feature of self-propelled aligning particles is their ability to spontaneously form an ordered state, \textit{i.e.} a flock. A large number of studies were dedicated to understand the nature of the flocking transition and how it is controlled by the density and noise in such systems~\cite{Ramaswamy2010, Marchetti2013}.

\begin{figure}
	\centering
	\includegraphics[width=.3\columnwidth]{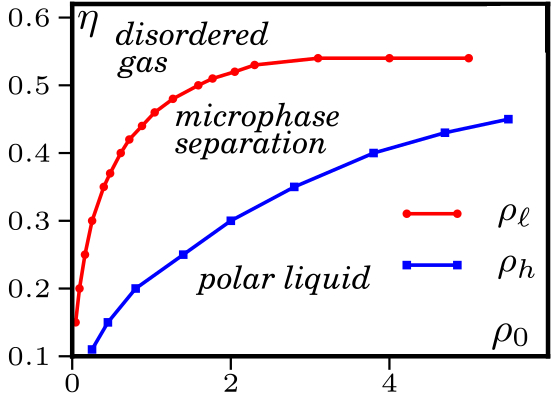}
	\hskip1cm
	\includegraphics[width=.22\columnwidth]{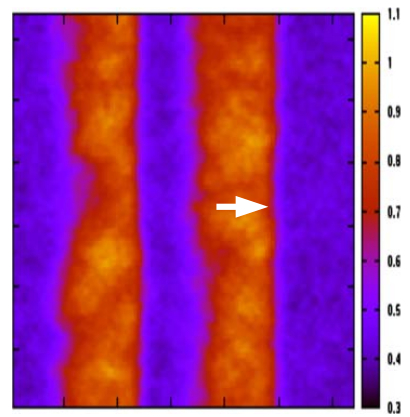}
	\caption{\label{fig:order}
		(Left) Phase diagram of the Vicsek model in the parameter space density-noise amplitude, respectively denoted by $\rho_0$ and $\eta$. Taken from~\cite{Solon2015b}. The coexistence regions between disordered and ordered phases are denoted by the binodal lines $\rho_\ell$ and $\rho_h$.
		(Right) Banding instability in the phase coexisting region of parameter space. The white arrow indicates the traveling direction. Taken from~\cite{Mishra2010}.
	}
\end{figure}

One of the first minimal models of flocking was proposed by analogy with ferromagnetism in two dimensions. The Vicsek model consists in describing the overdamped dynamics of $N$ point-like particles with fixed speed, whose alignment with their neighbors is set by some noisy rules in discrete time~\cite{Vicsek1995}. The order parameter of the system, defined as the mean velocity, quantifies the emergence of flocking at low noise and high density. It can be described as a first order transition akin to an equilibrium liquid-gas phase transition~\cite{Gregoire2004, Chate2008}, as illustrated in Fig.~\ref{fig:order}. The transition becomes continuous when alignment occurs through topological interactions, as reported in bird flocks~\cite{Ballerini2008, Cavagna2010, Ginelli2010}, or when adding birth and death reactions~\cite{Toner2012}.

The microscopic dynamics~\eqref{eq:dyn_inter} in the absence of steric repulsion and translational noise can be regarded as a continuous time formulation of the Vicsek model. To describe the properties of the system at large scale, the relevant hydrodynamics fields are the conserved particle density and the density of polarization. The polarization plays the dual role of mean velocity and order parameter of the system. Systematic perturbation methods have been employed to obtain some closed hydrodynamic equations~\cite{Bertin2006, Bertin2009, Peshkov2014}. Alternatively, continuum equations have been proposed from symmetry arguments. In their simplified form, the Toner-Tu equations can be written as~\cite{Toner1995, Toner1998}
\begin{equation}\label{eq:toner_tu}
	\p_t \rho = - v_0 \nabla \cdot {\bf p} , 
	\quad
	\p_t {\bf p} + \upsilon ( {\bf p} \cdot \nabla ) {\bf p} = - \brtg{ a + b \abs{ {\bf p} }^2 } {\bf p} + K \nabla^2 {\bf p} - \f{v_0}{2} \nabla \rho .
\end{equation}
The kinetic parameters can be expressed in terms of the microscopic ones. Considering a specific form of the interactions as ${\cal T} (\theta, {\bf r}) = \sin \theta / \pi R^2 $ when $ \abs{ {\bf r} } < R$, one gets $a = D_\text{r} - \rho \mu_\text{r} / 2$~\cite{Farrell2012}. The self-advection parameter $\upsilon$ being different from $v_0/\rho$, the hydrodynamic equations break the Galilean invariance in contrast to the Navier-Stokes equation. The terms $K \nabla^2 {\bf p}$ and $ - v_0 \nabla \rho / 2 $ are respectively analog to a viscous damping and a pressure gradient, and $ - \brtg{ a + b \abs{ {\bf p} }^2 } {\bf p} $ can be interpreted as a non-linear friction term. Alternatively, the left-hand side of~\eqref{eq:toner_tu} can be associated with the free-energy of a ferroelectric liquid crystal:
\begin{equation}
	{\cal F} = \int \brt{ \f{a}{2} {\bf p}^2 + \f{b}{4} {\bf p}^4 + \f{K}{2} (\p_\alpha p_\beta)^2 - \f{v_0 \rho}{2} \nabla \cdot {\bf p} } \dd \bf r .
\end{equation}
By construction, the Toner-Tu equations yield a continuous transition between a disordered state $ \abs {\bf p} = 0 $ for high noise and low density, and an ordered stated $ \abs{\bf p} = (-a/b)^{1/2} $ moving at constant velocity $v_0 \abs{\bf p}$ for low noise and high density. The transition line in the parameter space density-noise amplitude is set by the condition $a = 0$. The linear analysis reveals the existence of a banding instability along the direction of  spontaneously broken-symmetry~\cite{Bertin2009,Mishra2010, Solon2015b}, as reported in Fig.~\ref{fig:order}. An exact solutions can be found for the propagating profile in the traveling frame depending on the initial condition~\cite{Solon2015}. The breakdown of orientational symmetry leads to long-range interactions, in contrast with equilibrium, associated with some giant fluctuations of the number of particles in the ordered phase~\cite{Toner2005}. These fluctuations are quantified in numerical simulations from the structure factor at small wavenumber, and they have also been reported in experimental realizations of active ordered state~\cite{Narayan2007, Zhang2010}. To mimic the aligning interactions present in biological systems, systems of vibrated asymmetric particles have been designed as minimal experiments where the properties of the emerging order, either polar or nematic, can be quantified in great details~\cite{Kudrolli2008, Deseigne2010, Kumar2014}.

\subsection{Shape-driven interactions: density-independent rigidity and flocking transitions}

\begin{figure}
	\centering
	\includegraphics[width=.6\columnwidth]{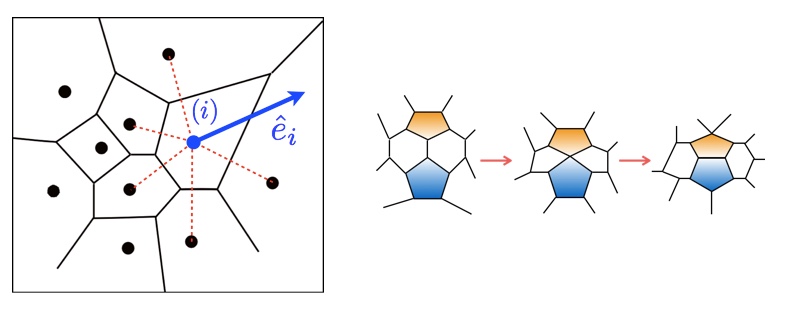}
	\caption{\label{fig:SPV}
		(Left) A sketch of the SPV model, showing the direction of instantaneous polarization $\hat{\mathbf{e}}_i$ of one cell. 
		(Right) Cells exchange neighbors via a $T_1$ transition.
	}
\end{figure}

The study of active jamming and active glasses formed by dense collections of ABPs was largely stimulated by the experimental observation of glassy dynamics in monolayer of epithelial cells~\cite{Angelini2010, Angelini2011, Nnetu2012, Puliafito2012}. On the other hand, in many developmental processes, such as embryogenesis, or in wound healing assays, epithelia are confluent, \textit{i.e.} they are densely packed sheets of cells that cover the plane, with no gaps between cells, and remain confluent while changing from liquid-like to solid-like behavior~\cite{Puliafito2012}. In other words, the packing fraction is always equal to unity, suggesting that, unlike in particulate systems, density may not be a relevant parameter to tune the onset of rigidity in epithelia. Recent work~\cite{Bi2014, Bi2015} has instead examined the materials properties of tissues by using active versions  of a well established model known as the Vertex Model (VM) that captures the interplay of cortical tension and cell-cell adhesion in controlling cellular arrangements in dense tissues~\cite{Nagai2001, Farhadifar2007, Hufnagel2007, Staple2010}. The VM describes a confluent epithelium as an array of densely packed columnar cells and  has been used successfully to reproduce experimental data in the \emph{Drosophila} embryo~\cite{Aigouy2010}. Neglecting cell height fluctuations, the VM describes a tissue as a two-dimensional  collection of irregular polygons representing the cell cross sections that tile the plane. Each cell is characterized by its area $A_i$ and perimeter $P_i$  and the tissue energy takes the form
\begin{equation}
\label{eq:Etissue}
E_\text{tissue}=\sum_{i=1}^N\left[\kappa_A(A_i-A_0)^2+\kappa_P(P_i-P_0)^2\right]\;.
\end{equation}
The first term in Eq.~\eqref{eq:Etissue} comes from bulk elasticity of the monolayer and its incompressibility in $3D$, with $A_0$ a cell target area. The second term describes the interplay between contractility in the actomyosin cortex and cell-cell adhesion, resulting in an effective boundary tension proportional to the target perimeter $P_0$. For $N$ cells in a square box of area $L^2$ with periodic boundary conditions, the average cell area $\bar A =L^2/N$ generally differs from $A_0$ and should be treated as an independent parameter. Since it does not affect the dynamics, here we choose $A_0 = \bar A$ and measure lengths in units of $A_0^{1/2}$ and energies in units of  $\kappa_A A_0^2$. The tissue energy is then controlled by two dimensionless parameters: the ratio $r=\kappa_A A_0/\kappa_P$ of the area and perimeter stiffnesses and a target shape index $s_0=P_0/A_0^{1/2}$ that describes cellular shape. The target shape index quantifies the anisotropy of polygonal shapes, with  $s_0 = 2 \pi^{1/2} \approx 3.544$ for a disk and $s_0 = 8^{1/2} 3^{1/4} \approx 3.722$ for a regular hexagon. By examining the disordered ground states of this energy, it was recently shown that this model exhibits a rigidity transition as a function of the shape parameter $s_0$~\cite{Bi2014,Bi2015}. The transition occurs at $s_0^*=3.81$, a value very close, but not equal, to that of a regular pentagon. For $s_0<s_0^*$, the system is rigid with glassy dynamics as there are large energy barriers for cellular rearrangements in the form of so-called $T_1$ transition shown in Fig.~\ref{fig:SPV}. 
For $s_0>s_0^*$ such energy barriers vanish and cells can freely rearrange, resulting in a liquid-like state. This work additionally identified the mean cellular shape $q=\langle P_i/A_i^{1/2}\rangle$ as an order parameter for the transition, where $\langle \cdot\rangle$ denotes an average over all cells. It is important to stress the distinction between the control parameter $s_0$, which describes single-cell properties and tunes the rigidity transition, and the observable $q(s_0)$, which measures the mean cellular shape in a given sample. The behavior of the mean cell shape $q$ as a function of the tuning parameter $s_0$ is analogous to that of magnetization as a function of decreasing temperature in a magnet: in the glassy solid $q(s_0)=3.81$ for $s_0<s_0^*$, while in the liquid it grows linearly with $s_0$. Hence, $q(s_0)$ can be regarded as an order parameter for the transition. This prediction has found experimental support in experiments on human bronchial epithelial cells~\cite{Park2016}. 
%
\begin{figure}[b!]
	\centering
	\includegraphics[width=.95\columnwidth]{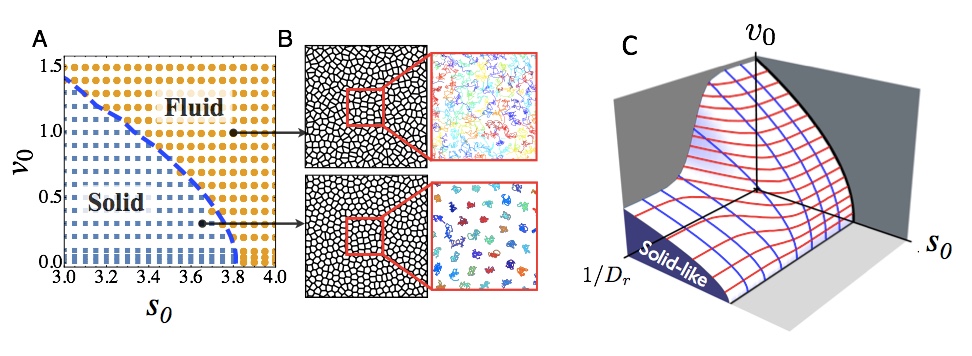}
	\caption{\label{fig:SPV_PD}
		(A) The phase diagram of the SPV model in the $(s_0,v_0)$ plane for fixed $D_\text{r}=1$ showing solid (blue squares) states where $D_\text{eff}$ vanishes and liquid (orange circles) states where $D_\text{eff}$ is finite.
		The blue dashed line is obtained by evaluating the mean cell shape $q$ and setting $q=3.81$, showing that dynamical and structural transitions coincide. 
		(B) Instantaneous tissue snapshots and cell tracks highlighting the different structure and dynamics in the two phases. 
		(C) The behavior of the tissue can be organize din a three-dimensional phase diagram as a function of motility $v_0$, persistence $1/D_\text{r}$ and target cell shape $s_0$.
		Taken from~\cite{Bi2016}.
	}
\end{figure}
On the other hand, cells are generally motile and out of equilibrium, pointing to the inadequacy of a description based on  energy minimization.  This observation motivated the formulation of a model of confluent epithelia that incorporates cell motility by  describing cells as self-propelled entities with the persistent dynamics of ABPs and interactions controlled by the tissue energy of the VM~\cite{Bi2016}.  Each cell is characterized by its position vector $\mathbf{r}_i$ and cell shape defined by the Voronoi tesselation of all cell positions (see Fig.~\ref{fig:SPV}). It is additionally endowed with motility described as a constant self-propulsion speed $v_0$ directed along the direction of cell polarization $\hat{\mathbf{e}}_i=(\cos\theta_i,\sin\theta_i)$, which as in ABPs  is in turn randomized by rotational noise. The dynamics is controlled by overdamped Langevin equations,
\begin{equation}
\label{eq:SPV}
\dot{\mathbf{r}}_i=-\mu_\text{t}\bm\nabla_{\mathbf{r}_i}E_\text{tissue}+v_0\hat{\mathbf{e}}_i\;,~~~~\dot{\theta_i}=(2D_\text{r})^{1/2}\eta_i\;,
\end{equation}
with $\eta_i(t)$ a white noise with unit variance as given in Eq.~\eqref{eq:noise}. We stress that, unlike in the particle systems considered earlier, the cell-cell interaction forces determined by the tissue energy  extend to second nearest neighbors. This model, dubbed Self-Propelled Voronoi model (SPV), has been studied numerically and shown to exhibit a liquid-solid transition tuned by cell shape $s_0$, cell motility $v_0$ and the persistence $\tau = 1 / D_\text{r}$ of single-cell dynamics~\cite{Bi2016}. The phase diagram in the $(s_0,v_0)$ plane is shown in Fig.~\ref{fig:SPV_PD}(a) and a rendering of the three-dimensional phase diagram is depicted in Fig.~\ref{fig:SPV_PD}(c).

As in ABPs, the  transition to the structurally arrested state was quantified by evaluating the cells' mean-square displacement and identified with the vanishing of the effective diffusivity defined 
as $D_\text{eff}=\lim_{t\rightarrow \infty}\langle\Delta\mathbf{r}^2(t)\rangle/4t$ (in practice the glass was defined as a system with $D_\text{eff}\leq 10^{-3}$). Remarkably, the same transition line is also obtained by evaluating the  structural order parameter $q$ and setting $q=3.81$. The identification of a glass transition in terms of both dynamical and structural order parameters is quite rare in glassy physics. This prediction can be verified experimentally by combining particle image velocimetry with cell shape data obtained form cell segmentation. Like in ABPs, increasing the persistence of single-cell dynamics promotes collective structural rearrangements that fluidify the tissue. The original work suggested that, in contrast to ABPs, where increasing persistence pushes the zero motility jamming transition to larger values of the packing fraction, in the SPV jamming at $v_0=0$ may actually occur always at $s_0=3.81$ for all values of $D_\text{r}$. It has now become apparent that earlier simulations may not have considered sufficiently low values of $v_0$ to reliably infer the $v_0\rightarrow 0$ behavior. In fact very recent work has indicated that, in contrast to the VM where the degrees of freedom are the vertices of the polygons, the highly constrained SPV that has  a much smaller number of degrees of freedom corresponding the cell positions may remain solid  for all values of $s_0$ at $v_0=0$~\cite{Sussman2017}. 
\begin{figure}
	\centering
	\includegraphics[width=.5\columnwidth]{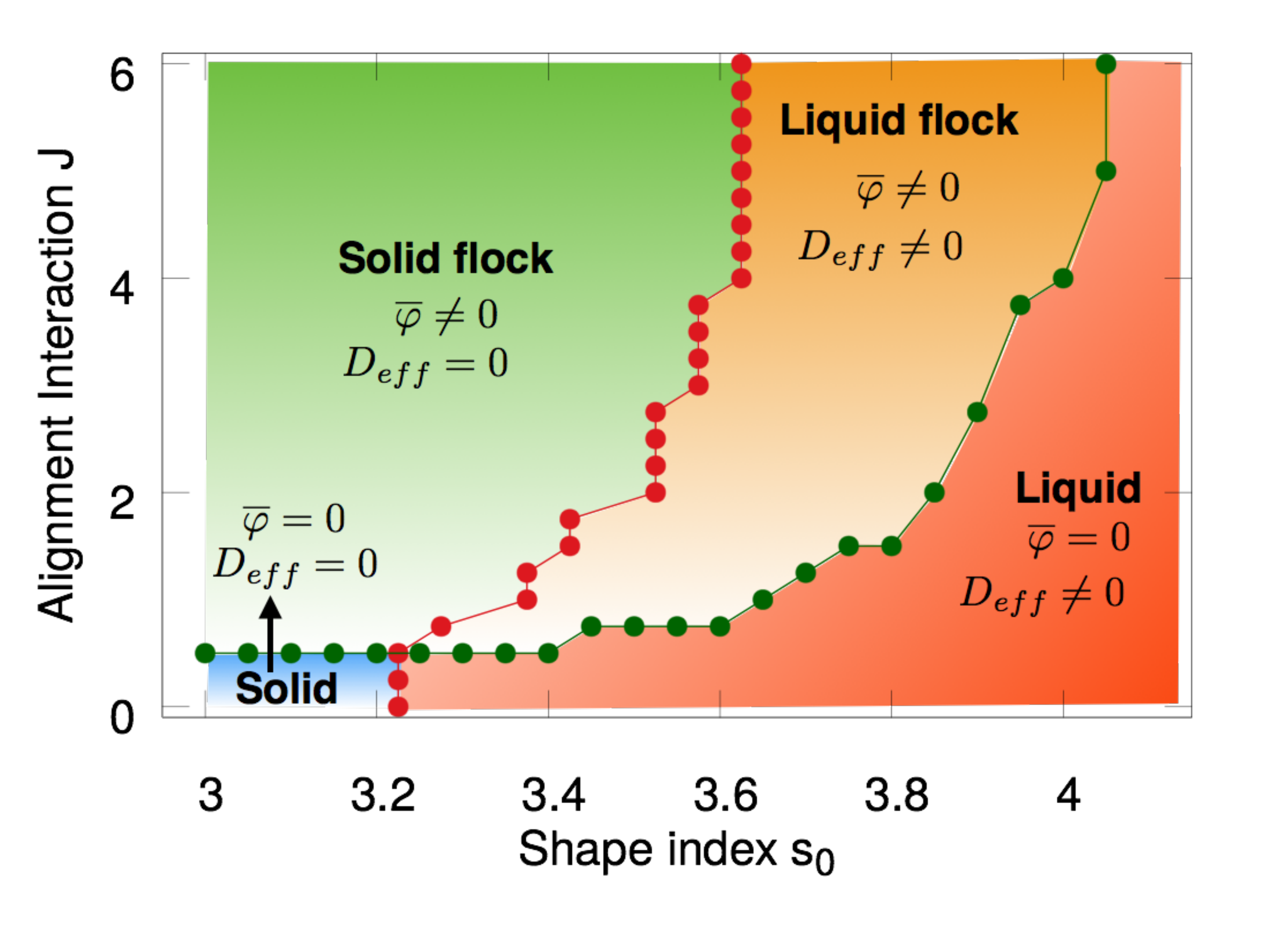}
	\caption{\label{fig:SPV_align}
 The phase diagram of the aligning SPV in the $(s_0,J)$ plane showing stationary and flocking solid and liquid states. Adapted from~\cite{Giavazzi2017}.
	}
\end{figure}
Cells in dense tissue often coordinate their direction of migration by aligning their local polarization~\cite{Ladoux2016}. This is evident for instance in wound healing assays where cells march coherently to fill empty space~\cite{Poujade2007}. Drawing again inspiration from active matter models of flocking, collective migration and large-scale cell streaming can be described by augmenting the SPV model to incorporate mechanical and biochemical stimuli from neighboring cells through an effective feedback mechanism at the single cell level that tends to align the polarization of each cell with the cell's own velocity, which is in turn controlled by interactions with other cells~\cite{Malinverno2017, Giavazzi2017}. This amount to replacing the equation governing the angular dynamics by
\begin{equation}
\dot\theta_i=J\sin(\phi_i-\theta_i)+(2D_\text{r})^{1/2}\eta_i\;,
\label{eq:align}
\end{equation}
where $\phi_i$ denotes the instantaneous direction of the cell velocity, $\mathbf{v}_i=|\mathbf{v}_i|(\cos\phi_i,\sin\phi_i)$. Note that this can also be interpreted as an interaction that aligns the cell polarization with the force exerted on the cell by the remainder of the tissue. This aligning interaction has been used before in particulate models~\cite{Szabo2006, Henkes2011}. This simple modification allows for the formation of flocking states, corresponding to situations where the cells  on average moving coherently in a direction selected spontaneously by the system. The transition can be quantified through a flocking order parameter, $\bar \varphi = N^{-1}\langle\big|(\sum_i\mathbf{v}_i/|\mathbf{v}_i|)\big|\rangle$.  At low alignment rate one obtains stationary jammed and liquid states  ($\bar \varphi =0$). As $J$ increases, one finds both a flocking solid ($D_\text{eff}=0$ and $\bar \varphi \neq 0$) and a flocking liquid ($D_\text{eff}\not=0$ and $\bar \varphi \not=0$), as shown in Fig.~\ref{fig:SPV_align}.

Alignment occurs provided the rate $J$ at which cells can coordinate their polarization exceeds the rate of structural rearrangements. In the solid where cells are caged and the only dynamics comes from orientational fluctuations, flocking  occurs as soon as $J$ exceeds $D_\text{r}$. Alignment additionally promotes solidification as evident from the slope of the line separating the flocking solid from the flocking liquid. This can be understood because alignment suppresses fluctuations transverse to the direction of mean motion, making it harder for cell to overcome the energy barriers for structural rearrangements associated with $T_1$ transitions. In the liquid, there is a region of parameters where the finite caging time exceeds the time for polarization alignment, resulting in a flocking liquid state. The flocking liquid is anisotropic, as revealed by the morphology of the dynamical heterogeneities observed in the system as the solid is approached from the liquid side. In the flocking state, these collective rearrangements resemble large scale collective streaming.

Numerics strongly indicate that the flocking transition in the SPV model is continuous with critical scaling, as found in Vicsek models with topological interactions~\cite{Ballerini2008,Cavagna2014} and in contrast to particle models with metric interactions where the transition has been established to be  first order~\cite{Gregoire2004}. Since in the SPV cells by definition always interact with their topological neighbors (albeit not just with the nearest neighbors), this suggest that the nature of the flocking transition may be intimately related to the topological versus metric nature of the interactions.


\section{Conclusions and outlook}

 We have presented a brief review of the rich behavior of collections of active particles that generate their own energy and spontaneously organize  in a variety of nonequilibrium states. The key property that distinguishes these systems from more familiar nonequilibrium systems is that active particles break time reversal symmetry (TRS) at the microscale, providing new, unexplored challenges to the statistical mechanician. The effects of the breaking of TRS, or, equivalently, the lack of detailed balance, become manifest in the presence of interparticle couplings or when active particles interact with walls or move in spatially or temporally inhomogeneous environments. We have focused here on the effect of interactions and highlighted a few key properties. Steric repulsions alone can suppress motility at high density resulting in condensed states even in the absence of any attractive interactions. Torques that tend to align the direction of motion of active agents yield transitions to flocking states where all agents move coherently in the absence of a leader and in the presence of noise. Interactions driven by shape changes that embody the interplay of cell contractility and cell-cell adhesion in dense tissues can tune density-independent transitions between liquid and arrested or solid-like states, that can either be stationary on average or exhibit mean flocking motion.

Our understanding of the behavior of active systems has proceeded at a very rapid pace in recent years. There remain, however, many challenges that may be grouped in three headings.

The first challenge is for theorists who are faced with  formulating the statistical mechanics of systems that break TRS at the level of the microdynamics. Recent work has shown that in active systems even the definition of state functions like temperature and pressure fails. A pressure equation of state has been constructed for ABPs~\cite{Solon2015c}, but it has also been shown that this is a special case and that in more general active systems where particles can exchange torques with each other or with the walls (because of their shape anisotropy or of explicit alignment interactions) the pressure depends on the properties of the wall~\cite{Solon2015d}. This failure of arguably one of the most familiar and basic notions of statistical physics highlights the need for new approaches for perhaps identifying effective state functions that incorporate the effect of the environment while remaining useful measures of the properties of the system. Work currently underway is attempting to develop a nonequilibrium thermodynamic of active systems and identify quantitative measures of their nonequilibrium nature~\cite{Nardini2016, Speck2016, Nardini2017, Seifert2017, Speck2017}. Also, still quite unexplored is the response of active system to additional external perturbations. Importantly, this theoretical work will need experimental validation, which is becoming available thanks to the development of ingenious man-made realization of active systems where controlled experiments are possible~\cite{Bricard2013, Bechinger2016}.

Secondly, we are presented with the challenge of harnessing active dynamics for the assembly of new materials with specific capabilities and functions. In mixture of active and passive particles, active agents can speed up the assembly of passive components by facilitating the overcome of energy barriers~\cite{Ni2014,Stenhammar2015}. Bacteria can transfer work to passive elements driving the motion of  mechanical components such as microgears~\cite{DiLeonardo2010,Sokolov2010}. Spatially and temporally inhomogeneous structures can be designed to control active dynamics and generate spontaneous flows and circulations~\cite{Stenhammar2016,Dogic2017}. But the application of these advances to the development of new materials requires the formulation of a systematic theoretical framework to quantify the design principles for active assembly -- something that is currently still lacking.

Finally, an important open question is whether ideas from active matter can have a true impact in biology. Active models, at both the continuum and mesoscopic levels, can describe the  transmission of mechanical forces that drives collective cell migration and tissue expansion in wound motility assays~\cite{Banerjee2015,Ladoux2016}. The recently developed models of rigidity transition in tissue seem to provide a promising framework for organizing biological data into phase diagrams where various mechanical and biochemical signals are described via a few effective parameters that characterize individual cellular properties, such as cell shape and  motility~\cite{Bi2016}. While this is encouraging, much more work remains to be done to provide a quantitative test to the theory.


\section*{Acknowledgements}

\noindent
We thank Adam Patch and Matteo Paoluzzi for help with some of the figures. MCM was supported at Syracuse University by the National Science Foundation through award DMR-1609208  the IGERT grant DGE-1068780 and by the Syracuse Soft Matter Program.


\appendix


\begin{thebibliography}{00}

\bibitem{Ramaswamy2010}
S. Ramaswamy, The mechanics and statistics of active matter, Annu. Rev. CMP
1, 323 (2010)

\bibitem{Marchetti2013}
M. C. Marchetti, J. F. Joanny, S. Ramaswamy, T. B. Liverpool, J. Prost, M. Rao, R. A. Simha, Hydrodynamics of soft active matter, Rev. Mod. Phys. 85, 1143 (2013)

\bibitem{Bechinger2016}
C. Bechinger, R. Di Leonardo, H. L\"owen, C. Reichhardt, G. Volpe, G. Volpe, Active particles in complex and crowded environments, Rev. Mod. Phys. 88, 045006 (2016)

\bibitem{Dombrowski2004}
C. Dombrowski, L. Cisneros, S. Chatkaew, R. E. Goldstein, J. O. Kessler, Self-concentration and large-scale coherence in bacterial dynamics, Phys. Rev. Lett. 93, 098103 (2004)

\bibitem{Sokolov2007}
A. Sokolov, I. S. Aranson, J. O. Kessler, R. E. Goldstein, Concentration dependence of the collective dynamics of swimming bacteria, Phys. Rev. Lett. 98, 158102 (2007)

\bibitem{Palacci2013}
J. Palacci, S. Sacanna, A. P. Steinberg, D. J. Pine, P. M. Chaikin, Living crystals of light-activated colloidal surfers, Science 339, 936 (2013)

\bibitem{Guo2014}
M. Guo, A. J. Ehrlicher, M. H. Jensen, M. Renz, J. R. Moore, R. D. Goldman, J. Lippincott-Schwartz, F. C. Mackintosh, D. A. Weitz, Probing the stochastic, motor-driven properties of the cytoplasm using force spectrum microscopy, Cell 158, 822 (2014)

\bibitem{Guo2015}
{\'E}. Fodor, M. Guo, N. S. Gov, P. Visco, D. A. Weitz, F. van Wijland, Activity-driven fluctuations in living cells, EPL 110, 48005 (2015)

\bibitem{Turlier2016}
H. Turlier, D. A. Fedosov, B. Audoly, T. Auth, N. S. Gov, C. Sykes, J.-F. Joanny, G. Gompper, T. Betz, Equilibrium physics breakdown reveals the active nature of red blood cell flickering, Nat. Phys. online (2016)

\bibitem{Ahmed2016}
{\'E}. Fodor, W. W. Ahmed, M. Almonacid, M. Bussonnier, N. S. Gov, M.-H. Verlhac, T. Betz, P. Visco, F. van Wijland, Nonequilibrium dissipation in living oocytes, EPL 116, 30008 (2016)

\bibitem{Gorfinkiel2011}
N. Gorfinkiel, G. B. Blanchard, Dynamics of actomyosin contractile activity during epithelial morphogenesis, Curr. Op. Cell Bio. 23, 531 (2011)

\bibitem{Guillot2013}
C. Guillot, T. Lecuit, Mechanics of epithelial tissue homeostasis and morphogenesis, Science 340, 1185 (2013)

\bibitem{Bi2016}
D. Bi, X. Yang, M. C. Marchetti, M. L. Manning, Motility-driven glass and jamming transitions in biological tissues, Phys. Rev. X 6, 021011 (2016)

\bibitem{Cavagna2014}
A. Cavagna, I. Giardina, Bird flocks as condensed matter, Annu. Rev. CMP 5,
183 (2014)

\bibitem{Vicsek1995}
T. Vicsek, A. Czirók, E. Ben-Jacob, I. Cohen, O. Shochet, Novel type of phase transition in a system of self-driven particles, Phys. Rev. Lett. 75, 1226 (1995)

\bibitem{Toner1995}
J. Toner, Y. Tu, Long-range order in a two-dimensional dynamical XY model: How birds fly together, Phys. Rev. Lett. 75, 4326 (1995)

\bibitem{Chate2006}
H. Chat\'e, F. Ginelli, R. Montagne, Simple model for active nematics: Quasi-long-range order and giant fluctuations, Phys. Rev. Lett. 96, 180602 (2006)

\bibitem{Schaller2010}
V. Schaller, C. Weber, C. Semmrich, E. Frey, A. R. Bausch, Polar patterns of driven filaments, Nature (London) 467, 73 (2010)

\bibitem{Fily2012}
Y. Fily, M. C. Marchetti, Athermal phase separation of self-propelled particles with no alignment, Phys. Rev. Lett. 108, 235702 (2012)

\bibitem{Cates2015}
M. E. Cates, J. Tailleur, Motility-induced phase separation, Annu. Rev. CMP 6, 219 (2015)

\bibitem{Perrin}
J. Perrin, L'agitation mol\'eculaire et le mouvement brownien, Comptes rendus hebdomadaires des s\'eances de l'Acad\'emie des sciences 146, 967 (1908)

\bibitem{Langevin}
P. Langevin, Sur la th\'eorie du mouvement brownien, Comptes rendus hebdomadaires des s\'eances de l'Acad\'emie des sciences 146, 530 (1908)

\bibitem{Kubo1966}
R. Kubo, The fluctuation-dissipation theorem, Rep. Prog. Phys. 29, 255 (1966)

\bibitem{Schnitzer1993}
M. J. Schnitzer, Theory of continuum random walks and application to chemotaxis, Phys. Rev. E 48, 2553 (1993)

\bibitem{Tailleur2008}
J. Tailleur, M. E. Cates, Statistical mechanics of interacting run-and-tumble bacteria, Phys. Rev. Lett. 100, 218103 (2008)

\bibitem{Hagan2013}
G. S. Redner, M. F. Hagan, A. Baskaran, Structure and dynamics of a phase-separating active colloidal fluid, Phys. Rev. Lett. 110, 055701 (2013)

\bibitem{Tailleur2013}
M. E. Cates, J. Tailleur, When are active brownian particles and run-and-tumble particles equivalent? Consequences for motility-induced phase separation, EPL 101, 20010 (2013)

\bibitem{Maggi2015}
C. Maggi, U. Marini Bettolo Marconi, N. Gnan, R. Di Leonardo, Multidimensional stationary probability distribution for interacting active particles, Sci. Rep. 5, 10742 (2015)

\bibitem{Farage2015}
T. F. F. Farage, P. Krinninger, J. M. Brader, Effective interactions in active brownian suspensions, Phys. Rev. E 91, 042310 (2015)

\bibitem{Nardini2016}
{\' E}. Fodor, C. Nardini, M. E. Cates, J. Tailleur, P. Visco, F. van Wijland, How far from equilibrium is active matter?, Phys. Rev. Lett. 117, 038103 (2016)

\bibitem{Szamel2014}
G. Szamel, Self-propelled particle in an external potential: Existence of an effective temperature, Phys. Rev. E 90, 012111 (2014)

\bibitem{Solon2015c}
A. P. Solon, M. Cates, J. Tailleur, Active brownian particles and run-and-tumble particles: A comparative study, EPJST 224, 1231 (2015)

\bibitem{Marconi2015}
U. Marini Bettolo Marconi, N. Gnan, C. Maggi, R. Di Leonardo, Velocity distribution in active particles systems, Sci. Rep. 6, 23297 (2015)

\bibitem{Simha2002}
R. Aditi Simha, S. Ramaswamy, Hydrodynamic fluctuations and instabilities in ordered suspensions of self-propelled particles, Phys. Rev. Lett. 89, 058101 (2002)

\bibitem{Martin1972}
P. C. Martin, O. Parodi, P. S. Pershan, Unified hydrodynamic theory for crystals, liquid crystals, and normal fluids, Phys. Rev. A 6, 2401 (1972)

\bibitem{Kruse2004}
K. Kruse, J. F. Joanny, F. J\"ulicher, J. Prost, K. Sekimoto, Asters, vortices, and rotating spirals in active gels of polar filaments, Phys. Rev. Lett. 92, 078101 (2004)

\bibitem{Bertin2006}
E. Bertin, M. Droz, G. Gr\'egoire, Boltzmann and hydrodynamic description for self-propelled particles, Phys. Rev. E 74, 022101 (2006)

\bibitem{Peshkov2014}
A. Peshkov, E. Bertin, F. Ginelli, H. Chat\'e, Boltzmann-ginzburg-landau approach for continuous descriptions of generic vicsek-like models, EPJE, 223, 1315 (2014)

\bibitem{Dean1996}
D. S. Dean, Langevin equation for the density of a system of interacting langevin processes, J. Phys. A: Math. Gen. 29, L613 (1996)

\bibitem{Buttinoni2013}
I. Buttinoni, J. Bialk\'e, F. K\"ummel, H. L\"owen, C. Bechinger, T. Speck, Dynamical clustering and phase separation in suspensions of self-propelled colloidal particles, Phys. Rev. Lett. 110, 238301 (2013)

\bibitem{Golestanian2012}
R. Golestanian, Collective behavior of thermally active colloids, Phys. Rev. Lett. 108, 038303 (2012)

\bibitem{Liebchen2015}
B. Liebchen, D. Marenduzzo, I. Pagonabarraga, M. E. Cates, Clustering and pattern formation in chemorepulsive active colloids, Phys. Rev. Lett. 115, 258301 (2015)

\bibitem{Cates2010}
M. E. Cates, D. Marenduzzo, I. Pagonabarraga, J. Tailleur, Arrested phase separation in reproducing bacteria creates a generic route to pattern formation, Proc. Natl. Acad. Sci. USA 107, 11715 (2010)

\bibitem{Yang2014}
X. Yang, D. Marenduzzo, M. C. Marchetti, Spiral and never-settling patterns in active systems, Phys. Rev. E 89, 012711 (2014)

\bibitem{Bialke2013}
J. Bialk\'e, H. L\"owen, T. Speck, Microscopic theory for the phase separation of self-propelled repulsive disks, EPL 103, 30008 (2013)

\bibitem{Fily2014}
Y. Fily, S. Henkes, M. C. Marchetti, Freezing and phase separation of self-propelled disks, Soft Matter 10, 2132 (2014)

\bibitem{Solon2016}
A. P. Solon, J. Stenhammar, M. E. Cates, Y. Kafri, J. Tailleur, Generalized thermodynamics of phase equilibria in scalar active matter, arXiv:1609.03483

\bibitem{Wittkowski2014}
R. Wittkowski, A. Tiribocchi, J. Stenhammar, R. J. Allen, D. Marenduzzo, M. E. Cates, Scalar $\phi^4$ field theory for active-particle phase separation, Nat. Comm. 5, 4351 (2014)

\bibitem{Tiribocchi2015}
A. Tiribocchi, R. Wittkowski, D. Marenduzzo, M. E. Cates, Active model H: Scalar active matter in a momentum-conserving fluid, Phys. Rev. Lett. 115, 188302 (2015)

\bibitem{Nardini2017}
C. Nardini, \'E. Fodor, E. Tjhung, F. van Wijland, J. Tailleur, M. E. Cates, Entropy production in field theories without time-reversal symmetry: Quantifying the non-equilibrium character of active matter, Phys. Rev. X 7, 021007 (2017)

\bibitem{Reichhardt2014}
C. Reichhardt, C. J. Olson Reichhardt, Absorbing phase transitions and dynamic freezing in running active matter systems, Soft Matter 10, 7502 (2014)

\bibitem{Bialke2012}
J. Bialk\'e, T. Speck, H. L\"owen, Crystallization in a dense suspension of self-propelled particles, Phys. Rev. Lett. 108, 168301 (2012)

\bibitem{Menzel2014}
A. M. Menzel, T. Ohta, H. L\"owen, Active crystals and their stability, Phys. Rev. E 89 022301 (2014)

\bibitem{Berthier2014}
L. Berthier, Nonequilibrium glassy dynamics of self-propelled hard disks, Phys.
 Rev. Lett. 112, 220602 (2014) 

\bibitem{Lu2013}
P. J. Lu, D. A. Weitz, Colloidal particles: crystals, glasses, and gels, Annu. Rev. CMP 4, 217 (2013)

\bibitem{Levis2015}
D. Levis, L. Berthier, From single-particle to collective effective temperatures in an active fluid of self-propelled particles, EPL 111, 60006 (2015)

\bibitem{Ni2014}
R. Ni, M. A. Cohen Stuart, M. Dijkstra, P.G. Bolhuis, Crystallizing hard-sphere glasses by doping with active particles, Soft Matter 10, 6609 (2014)

\bibitem{Ni2014b}
R. Ni, M. A. Cohen Stuart, M. Dijkstra, Pushing the glass transition towards random close packing using self-propelled hard spheres, Nat. Commun. 4, 2704 (2014)

\bibitem{Szamel2015}
G. Szamel, E. Flenner, L. Berthier, Glassy dynamics of athermal self-propelled particles: Computer simulations and a nonequilibrium microscopic theory, Phys. Rev. E 91, 062304 (2015)

\bibitem{Flenner2016}
E. Flenner, G. Szamel, L. Berthier, The nonequilibrium glassy dynamics of self-propelled particles, Soft Matter 12, 7136 (2016)

\bibitem{Nandi2017}
S. Kumar Nandi, N. S. Gov, Nonequilibrium mode-coupling theory for dense active systems of self-propelled particles, arXiv:1708.05222

\bibitem{Narayan2007}
V. Narayan, S. Ramaswamy, N. Menon, Long-lived giant number fluctuations in a swarming granular nematic, Science 317, 105 (2007)

\bibitem{Bialek2012}
W. Bialek, A. Cavagna, I. Giardina, T. Mora, E. Silvestri, M. Viale, A. M. Walczak, Statistical mechanics for natural flocks of birds, Proc. Natl. Acad. Sci. USA 109, 4786 (2012)

\bibitem{Gregoire2004}
Guillaume Gr\'egoire, Hugues Chat\'e, Onset of collective and cohesive motion, Phys. Rev. Lett. 92, 025702 (2004)

\bibitem{Chate2008}
Hugues Chat\'e, Francesco Ginelli, Guillaume Gr\'egoire, Franck Raynaud, Collective motion of self-propelled particles interacting without cohesion, Phys. Rev. E 77, 046113 (2008)

\bibitem{Ginelli2010}
Francesco Ginelli, Hugues Chat\'e, Relevance of metric-free interactions in flocking phenomena, Phys. Rev. Lett. 105, 168103 (2010)

\bibitem{Ballerini2008}
M. Ballerini, N. Cabibbo, R. Candelier, A. Cavagna, E. Cisbani, I. Giardina, V. Lecomte, A. Orlandi, G. Parisi, A. Procaccini, M. Viale, V. Zdravkovic, Interaction ruling animal collective behavior depends on topological rather than metric distance: Evidence from a field study, Proc. Natl. Acad. Sci. USA 105, 1232 (2008)

\bibitem{Cavagna2010}
A. Cavagna, A. Cimarelli, I. Giardina, G. Parisi, R. Santagati, F. Stefanini, M. Viale, Scale-free correlations in starling flocks, Proc. Natl. Acad. Sci. USA 107, 11865 (2010)

\bibitem{Toner2012}
J. Toner, Birth, death, and flight: A theory of malthusian flocks, Phys. Rev. Lett. 108, 088102 (2012)

\bibitem{Bertin2009}
E. Bertin, M. Droz, G. Gr\'egoire, Hydrodynamic equations for self-propelled particles: microscopic derivation and stability analysis, J. Phys. A: Math.
Theor. 42, 445001 (2009)

\bibitem{Toner1998}
J. Toner, Y. Tu, Flocks, herds, and schools: A quantitative theory of flocking, Phys. Rev. E 58, 4828 (1998)

\bibitem{Farrell2012}
F. D. C. Farrell, M. C. Marchetti, D. Marenduzzo, J. Tailleur, Pattern formation in self-propelled particles with density-dependent motility, 
Phys. Rev. Lett. 108, 248101 (2012)

\bibitem{Mishra2010}
S. Mishra, A. Baskaran, M. C. Marchetti, Fluctuations and pattern formation in self-propelled particles, Phys. Rev. E 81, 061916 (2010)

\bibitem{Solon2015b}
A. P. Solon, H. Chat\'e, J. Tailleur, From phase to microphase separation in flocking models: The essential role of nonequilibrium fluctuations, Phys. Rev. Lett. 114, 068101 (2015)

\bibitem{Solon2015}
A. P. Solon, J.-B. Caussin, D. Bartolo, H. Chat\'e, J. Tailleur, Pattern formation in flocking models: A hydrodynamic description, Phys. Rev. E 92, 062111 (2015)

\bibitem{Toner2005}
J. Toner, Y. Tu, S. Ramaswamy, Hydrodynamics and phases of flocks, Ann. Phys. (Amsterdam) 318, 170 (2005)

\bibitem{Zhang2010}
H. P. Zhang, A. Be’er, E.-L. Florin, H. L. Swinney, Collective motion and density fluctuations in bacterial colonies, Proc. Natl. Acad. Sci. USA 107,  13626 (2010)

\bibitem{Kudrolli2008}
Arshad Kudrolli, Geoffroy Lumay, Dmitri Volfson, Lev S. Tsimring, Swarming and swirling in self-propelled polar granular rods, Phys. Rev. Lett. 100, 058001 (2008)

\bibitem{Deseigne2010}
Julien Deseigne, Olivier Dauchot, Hugues Chat\'e, Collective motion of vibrated polar disk, Phys. Rev. Lett. 105, 098001 (2010)

\bibitem{Kumar2014}
Nitin Kumar, Harsh Soni, Sriram Ramaswamy, A.K. Sood, Flocking at a distance in active granular matter, Nat. Com. 5, 4688 (2014)

\bibitem{Angelini2010}
T. E. Angelini, E. Hannezo, X. Trepat, J. J. Fredberg, D. A. Weitz, Cell migration driven by cooperative substrate deformation patterns, Phys. Rev. Lett. 104, 168104 (2010)

\bibitem{Angelini2011}
T. E. Angelini, E. Hannezo, X. Trepat, M. Marquez, J. J. Fredberg, and D. A. Weitz, Glass-like dynamics of collective cell migration, Proc. Natl. Acad. Sci. USA 108, 4714 (2011)

\bibitem{Nnetu2012}
K. D. Nnetu, M. Knorr, J. K\"as, M. Zink, The impact of jamming on boundaries of collectively moving weak-interacting cells, New J. Phys. 14, 115012 (2012)

\bibitem{Puliafito2012}
A. Puliafito, L. Hufnagel, P. Neveu, S. Streichan, A. Sigal, D. K. Fygenson, B. I. Shraiman, Collective and single cell behavior in epithelial contact inhibition, Proc. Natl. Acad. Sci. USA 109, 739 (2012)
 
\bibitem{Bi2014}
D. Bi, J. H. Lopez, J. M. Schwarz, M. L. Manning, Energy barriers and cell migration in densely packed tissues, Soft Matter 10, 1885 (2014)

\bibitem{Bi2015}
D. Bi, J. H. Lopez, J. M. Schwarz, M. L. Manning, A density-independent rigidity transition in biological tissues, Nat. Phys. 11, 1074 (2015)
 
\bibitem{Nagai2001}
T. Nagai, H. Honda, A dynamic cell model for the formation of epithelial tissues, Philos. Mag. B 81, 699 (2001)

\bibitem{Farhadifar2007}
R. Farhadifar, J.-C. Roeper, B. Aigouy, S. Eaton, F. J\"ulicher, The influence of cell mechanics, cell-cell interactions, and proliferation on epithelial packing, Curr. Biol. 17, 2095 (2007)
 
\bibitem{Hufnagel2007}
L. Hufnagel, A. A. Teleman, H. Rouault, S. M. Cohen, B. I. Shraiman, On the mechanism of wing size determination in fly development, Proc. Natl. Acad. Sci. USA 104, 3835 (2007)
 
\bibitem{Staple2010}
D. B. Staple, R. Farhadifar, J. C. Roeper, B. Aigouy, S. Eaton, F. J\"ulicher, Mechanics and remodelling of cell packings in epithelia, EPJE 33, 117 (2010)

\bibitem{Aigouy2010}
B. Aigouy, R. Farhadifar, D. B. Staple, A. Sagner, J. C. R\"oper, F. J\"ulicher, S. Eaton, Cell flow reorients the axis of planar polarity in the wing epithelium of drosophila, Cell 142, 773 (2010)

\bibitem{Park2016}
J.-A. Park, J. Hun Kim, D. Bi, J. A. Mitchel, N. Taheri Qazvini, K. Tantisira,	C. Young Park, M. McGill, S.-H. Kim, B. Gweon, J. Notbohm, R. Steward Jr,	S. Burger, S. H. Randell, A. T. Kho,	D. T. Tambe, C. Hardin,	S. A. Shore, E. Israel, D. A. Weitz, D. J. Tschumperlin, E. P. Henske, S. T. Weiss,	M. L. Manning, J. P. Butler, J. M. Drazen, J. J. Fredberg, Unjamming and cell shape in the asthmatic airway epithelium, Nat. Mater. 14, 1040 (2015)

\bibitem{Sussman2017}
D. M. Sussman, M. Merkel, No unjamming transition in a marginal vertex model of biological tissue, arXiv:1708.03396

\bibitem{Ladoux2016}
B. Ladoux, R.-M. M\`ege, X. Trepat, Front-rear polarization by mechanical cues: from single cells to tissues, Trends Cell Biol. 26, 420 (2016)

\bibitem{Poujade2007}
M. Poujade, E. Grasland-Mongrain, A. Hertzog, J. Jouanneau, P. Chavrier, B. Ladoux, A. Buguin, P. Silberzan, Collective migration of an epithelial monolayer in response to a model wound, Proc. Nat. Acad. Sci. USA 104, 15988 (2007)

\bibitem{Malinverno2017}
C. Malinverno, S. Corallino, F. Giavazzi,	M. Bergert,	Q. Li, M. Leoni, A. Disanza, E. Frittoli,	A. Oldani, E. Martini, T. Lendenmann,	G. Deflorian,	G. V. Beznoussenko,	D. Poulikakos, K. Haur Ong,	M. Uroz, X. Trepat,	D. Parazzoli,	P. Maiuri,	W. Yu, A. Ferrari, R. Cerbino, G. Scita, Endocytic reawakening of motility in jammed epithelia, Nat. Mater. 16, 587 (2017) 

\bibitem{Giavazzi2017}
F. Giavazzi, M. Paoluzzi, M. Macchia, D. Bi, G. Scita, M. L. Manning, R. Cerbino, M. C. Marchetti, Flocking transition in confluent tissues, arXiv:1706.01113

\bibitem{Szabo2006}
B. Szab\'o, G. J. Sz\"oll\"osi, B. G\"onci, Zs. Jur\'anyi, D. Selmeczi, T. Vicsek, Phase transition in the collective migration of tissue cells: Experiment and model, Phys. Rev. E 74, 061908 (2006)

\bibitem{Henkes2011}
Silke Henkes, Yaouen Fily, M. Cristina Marchetti, Active jamming: Self-propelled soft particles at high density, Phys. Rev. E 84, 040301(R) (2011)

\bibitem{Solon2015c}
A. P. Solon, J. Stenhammar, R. Wittkowski, M. Kardar, Y. Kafri, M. E. Cates, J. Tailleur, Pressure and phase equilibria in interacting active brownian spheres,
Phys. Rev. Lett. 114, 198301 (2015)

\bibitem{Solon2015d}
A. P. Solon, Y. Fily, A. Baskaran, M. E. Cates, Y. Kafri, M. Kardar, J. Tailleur, Pressure is not a state function for generic active fluids,
Nat. Phys. online (2015)

\bibitem{Speck2016}
Thomas Speck, Stochastic thermodynamics for active matter, EPL 114, 30006 (2016)

\bibitem{Seifert2017}
Patrick Pietzonka, Udo Seifert, Entropy production of active particles and for particles in active baths, J. Phys. A: Math. Theor. 51, 01LT01 (2017)

\bibitem{Speck2017}
Thomas Speck, Stochastic thermodynamics with reservoirs:  Sheared and active colloidal particles, arXiv:1707.05289

\bibitem{Bricard2013} 
A. Bricard, J.-B. Caussin, N. Desreumaux, O. Dauchot, D. Bartolo,  Emergence of macroscopic directed motion in populations of motile colloids, Nature 503, 95 (2013)
 
\bibitem{Stenhammar2015} 
J. Stenhammar, R. Wittkowski, D. Marenduzzo, M. E. Cates, Activity-induced phase separation and self-assembly in mixtures of active and passive particles, Phys. Rev. Lett. 114, 018301 (2015) 

\bibitem{DiLeonardo2010}
R. Di Leonardo, L. Angelani, D. Dell'Arciprete, G. Ruocco, V. Iebba, S. Schippac, M. P. Conte, F. Mecarini, F. De Angelis, E. Di Fabrizio, Bacterial ratchet motors, Proc. Natl. Acad. Sci. USA 107, 9541 (2010)

\bibitem{Sokolov2010}
A. Sokolov, M. M. Apadoca, B. A. Grzybowski, I. S. Aranson, Swimming bacteria power microscopic gears, Proc. Natl. Acad. Sci. USA 107, 969 (2010)

\bibitem{Stenhammar2016} 
J. Stenhammar, R. Wittkowski, D. Marenduzzo, M. E. Cates, Light-induced self-assembly of active rectification devices,  Sci. Adv. 2, e1501850 (2016) 

\bibitem{Dogic2017}
K.-T. Wu, J. B. Hishamunda, D. T. N. Chen, S. J. DeCamp, Y.-W. Chang, A. Fern\'andez-Nieves, S. Fraden, Z. Dogic, Transition from turbulent to coherent flows in confined three-dimensional active fluids, Science 355, 6331 (2017)

\bibitem{Banerjee2015}
S. Banerjee, K. J. C. Utuje, M. C. Marchetti, Propagating stress waves during epithelial expansion, Phys. Rev. Lett. 114, 228101 (2015)


\end{thebibliography}
\end{document}